\documentclass{aa}

\usepackage{txfonts}
\usepackage{natbib}
\usepackage{graphicx}

\bibpunct{(}{)}{;}{a}{}{,} 

\begin{document}

\title{Dust Extinction and Emission in a Clumpy Galactic Disk}
\subtitle{An Application of the Radiative Transfer Code TRADING}

\author{Simone Bianchi}
\institute{INAF-Istituto di Radioastronomia, Sezione di Firenze,  
           Largo Enrico Fermi 5, 50125 Firenze, Italy \\
           \email{sbianchi@arcetri.astro.it}
          }

\date{Received 22 April 2008 / Accepted 18 July 2008}

\abstract
{}
{I present the Monte Carlo radiative transfer code TRADING 
(Transfer of RAdiation through Dust In Galaxies). The code computes 
self-consistently the extinction of radiation in a dusty medium
(including absorption and scattering) and the dust emission.}
{A binary-tree adaptive grid is used for the description of the 
dust distribution. Dust radiation is computed at thermal equilibrium 
or under stochastic heating condition, for a distribution of grains of
different radii and materials. The code is applied to the case of a
clumpy galactic disk, including both diffuse dust and a distribution 
of spherical clouds modelled on the GMCs of local galaxies. Diffuse
and localised sources of starlight are used, with independent spectra.}
{A model is provided for the edge-on galaxy NGC~891. The SED of the 
galaxy from the UV to the submm/mm range can be well reproduced by: 
a bulge/disk configuration of old stars together with an extended 
dust disk, as suggested by the analysis of optical/near-infrared images;
a clumpy dust distribution of the same mass as the diffuse dust disk,
together with a UV emitting component, half of which in the form of a
diffuse disk and half in sources embedded in clouds. In total, 
it is found that about 35\% of the bolometric radiation is absorbed 
(and emitted) by dust; and that absorption of starlight from the old 
population contributes to about 60\% of the dust emission. 
A significant component of the dust emission from clouds is due 
to absorption of diffuse radiation. Radial profiles of dust emission
in a clumpy disk are almost independent of the wavelength, with the
exception of the wavelength range on the Wien side of the thermal 
equilibrium peak.
}
{}


\maketitle

\section{Introduction}

The last two decades have seen a good effort in studying the role of 
dust in the appearance of a spiral galaxy\footnote{A parallel and
more considerable development has occurred in radiative transfer studies
of dusty circumstellar environments \citep[see, e.g., the various 
models described in ][]{PascucciA&A2004}.}. Radiative transfer models 
have shown that it is necessary to include scattering and consider 
appropriate geometries for a proper analysis of the internal extinction 
\citep[][to cite only the most recent works]{PieriniApJ2004,TuffsA&A2004,
RochaMNRAS2008}. 
Building on models of extinction in the optical, simulations of dust 
emission in the Far-Infrared (FIR) have been produced
\citep{BianchiA&A2000b,PopescuA&A2000,BaesProc2004}.

Puzzling has been the comparison between models and observations.
Dust disks of moderate optical depth can reproduce observations
of edge-on spirals \citep{XilourisSub1998,BianchiA&A2007}, in accordance
with the analysis of extinction of  background sources by less inclined
disks \citep[see, e.g., ][]{HolwerdaAJ2007}. Instead, the Spectral
Energy Distribution (SED) in FIR/submm requires dust absorption
and emission from a significantly larger mass 
\citep{BianchiA&A2000b,PopescuA&A2000,MisiriotisA&A2001}.

The {\em excess} mass derived from emission models could be 
associated with molecular clouds \citep{PopescuA&A2000}. Thus,
it becomes necessary to study the influence of inhomogeneities,
or {\em clumping}, on the radiative transfer. So far, the effects
of clumping have been studied mainly in extinction for galactic disks 
\citep{BianchiSub1999,MatthewsApJ2001,PieriniApJ2004} and in emission 
for geometries appropriate to starburst galaxies \citep{MisseltApJ2001}.
In all these cases, clumps have been modelled as higher density cells
in a uniformly spaced dust grid. However, the lack of resolution inside 
a clumpy cell does not allow to study cases in which significant
emission comes from within the cell, as when modelling young
stellar sources embedded in a molecular cloud. To obviate this,
\citet{SilvaApJprep1998} use a complete radiative transfer solution for
a single cloud, and add the cumulative output from clouds to the dust 
emission from the diffuse medium. A more self-consistent solution
would require the usage of an adaptive mesh refinement grid, with
a finer resolution for the denser medium.

I present in this paper the Monte Carlo (MC) radiative transfer code
TRADING (Transfer of RAdiation through Dust IN Galaxies\footnote{The 
acronym was suggested by a misprint in a notice at the {\em Exploratory 
Workshop:Tracing Dust In Spiral Galaxies}. The workshop was funded by 
the European Science Foundation and took place in Ghent, Belgium on 
May 13-17, 2007.}). For selected viewing directions, the code produces 
images of dust-extinguished starlight in the Ultraviolet (UV), Optical 
and Near Infrared (NIR), and of dust emission from the Mid-Infrared 
(MIR) to the submm. Images can be integrated to obtain the total SED. 
TRADING builds on our previous models and has been developed
from the regular grid version used in \citet{BianchiA&A2007}.

The scheme of TRADING is similar to that of the code DIRTY 
\citep{GordonApJ2001,MisseltApJ2001}. However, in TRADING the dust 
distribution is described with an adaptive grid, while DIRTY is based 
on a regular grid.
An adaptive grid has been included in the dust radiative transfer codes 
SUNRISE~\citep{JonssonMNRAS2006}, which has been used to study extinction 
in hydrodynamical simulations of isolated spirals \citep{RochaMNRAS2008},
and ART$^2$ \citep{LiApJ2008}. SUNRISE does not simulate dust emission, 
while ART$^2$ includes thermal equilibrium heating from a single grain. 
In TRADING, instead, emission is fully taken into account by using a 
distribution of grain sizes and materials; it includes both thermal and 
stochastic heating, and self-absorption.
With DIRTY, SUNRISE and ART$^2$ (and many other codes in the radiative
transfer literature) TRADING shares the use of the MC technique,
as this is of easy implementation in any geometrical configuration and
not necessarily heavier in computing time than analytical techniques
\citep{BaesMNRAS2001a}.

TRADING has been developed, and used in this paper, to study the 
influence of inhomogeneities (clouds) in the dust emission from a
galactic disk. 
To my knowledge, this is the first attempt ever to model the influence 
of dust on a galactic SED by using a fully self consistent radiative 
transfer calculation over scales ranging from molecular clouds
to the galactic disk.
Despite the code is admittedly tuned to this problem, however,
it is of general use and can be easily adapted to other 
astrophysical scenarios.

The scheme of the paper is as follows. The code is described in
Sect.~\ref{code}. In Sect.~\ref{dust}, I present the implementation of 
the \citet{DraineApJ2007b} dust model used in this paper.
In Sect.~\ref{results} TRADING is used to model the global SED of 
NGC891. The results of this application, together with the main
features of TRADING, are summarised in Sect.~\ref{summary}.

\section{The code}
\label{code}

TRADING consists of two main parts: the first follows the MC random 
walk of stellar energy packets ({\em photons}) in the dusty medium, through 
absorptions and scattering; a 3-D map of the Interstellar Radiation Field 
(ISRF) is produced and fed to the second part of the code, which computes 
dust emission.

The application of the MC technique to radiative transfer problems is
well described in the literature. Thus, I do not provide here full details
on all the algorithms used in TRADING but I will refer the reader to other 
published works \citep[in particular, to ][]{BianchiApJ1996,GordonApJ2001,
MisseltApJ2001,BaesMNRAS2003,NiccoliniA&A2003,JonssonMNRAS2006}.

\subsection{Stellar Emission}
\label{stellar_emission}

In TRADING, stellar radiation can be emitted from three different kind 
of distributions: an exponential disk, a spheroidal bulge, a collection of 
discrete spherical or point-like sources. 

As usual in models of spiral galaxies, the luminosity density of 
the stellar disk is given by an exponential distribution along the
radial, $r=\sqrt{x^2+y^2}$, and vertical, $z$, coordinates. It is
\begin{equation}
\rho^\mathrm{disk}(r,z)=\rho^\mathrm{disk}_0 
  \exp\left[-\frac{r}{h_\mathrm{s}} -\frac{|z|}{z_\mathrm{s}}\right],
\label{stardisk}
\end{equation}
where  $h_\mathrm{s}$ and $z_\mathrm{s}$ are the radial and vertical
scalelength of the stellar distribution, and $\rho^\mathrm{disk}_0$
is the central luminosity density.

The de-projection of a \citet{SersicBook1968} profile is used for the 
spheroidal bulge. For a profile index $n$, the luminosity density
can be written as
\begin{equation}
\rho^\mathrm{bulge}(r,z)= \rho^\mathrm{bulge}_1
\frac{\exp[b_n(1-B^{1/n})]}{B^\alpha},
\label{sersic}
\end{equation}
where $\rho^\mathrm{bulge}_1$ is the luminosity density at $B=1$,
\[
B=\frac{\sqrt{r^2+z^2/(b/a)^2}}{R_\mathrm{e}},
\]
\[
b_n=2n-1/3-0.009876/n, \qquad \alpha=(2n-1)/2n,
\]
$b/a$ is the minor/major axis ratio and $R_\mathrm{e}$ the effective radius
\citep{PrugnielA&A1997}. Bulges with $n=1,2,3$ or 4 are implemented in TRADING.
For $n=4$ the projection of Eq.~\ref{sersic} on the sky plane is the 
$R^{1/4}$ \citep{DeVaucouleursBook1959} surface brightness profile.

To simulate emission from star-forming regions within GMCs 
(Sect.~\ref{dust_distribution}), TRADING allows radiation to be emitted
also from a list of spherical sources. Each source is characterised by its
fractional luminosity relative to the total luminosity of all sources
in the list, $\ell_i$; by its radius, $r_i$; and by its center in space. 
In the present version of TRADING, the luminosity density within each 
source is constant (though I mostly use in this paper 
point-like sources, $r_i=0$). This can be easily changed to account 
for a specific radial distribution. For example, the SPH smoothing kernel 
could be used to simulate emission from sources in hydrodynamical 
simulations \citep{JonssonMNRAS2006}.

Several emission components can be included in a simulation. Each component
is described by its set of parameters and by its fractional luminosity $l_i$
with respect to the total model luminosity. The position of emission of a 
{\em photon} is then derived using the MC method. First, a component $i$ is 
selected so that
\begin{equation}
\sum_{k=0}^{i} l_k < \mathcal{R} \le \sum_{k=0}^{i+1} l_k. 
\label{lumsel}
\end{equation}
Throughout this paper, $\mathcal{R}$ is a random number uniformly
distributed in $[0,1]$.  Then, the position within the selected
component is derived: the procedures to obtain the $x$,$y$,$z$
coordinates of {\em photon} emission from the exponential disk and the bulge
are described in \citet{BianchiApJ1996} and \citet{BaesMNRAS2003}. To 
save computing time, a table with $L^\mathrm{bulge}(r)$, the luminosity
of a Sersic bulge emitted within radius $r$, is provided to the code, 
and the distance from the bulge center is obtained through interpolation.

If the {\em photon} is emitted within the list of spherical sources, the 
specific source first need to be selected. This is achieved by applying 
Eq.~\ref{lumsel} to the fractional luminosity of a source $\ell_i$. 
The  $x$,$y$,$z$ coordinates are then extracted in a manner analogous to
the bulge, with the radial distance from the source center given by 
$r_i \mathcal{R}^{1/3}$, as results from applying the MC method to a 
homogeneous sphere.

The direction of emission of {\em photons} is isotropic in the solid angle, 
and the initial {\em photon} energy (or {\em weight}) is set to unity. 
The {\em photon} wavelength $\lambda$ is obtained from the adopted stellar 
spectrum $F_{\lambda}$,
\begin{equation}
\int_{\lambda^\mathrm{stars}_\mathrm{min}}^{\lambda}
F_{\lambda} d\lambda= \mathcal{R} 
\int_{\lambda^\mathrm{stars}_\mathrm{min}}^{\lambda^\mathrm{stars}_\mathrm{max}}
F_{\lambda} d\lambda,
\label{mcspectrum}
\end{equation}
with $[\lambda^\mathrm{stars}_\mathrm{min},\lambda^\mathrm{stars}_\mathrm{max}]$
the wavelength range for stellar emission. To avoid 
numerical inversion of Eq.~\ref{mcspectrum} within the code, a table
is provided for each spectrum. Each of the emission components can have an 
independent stellar spectrum.

\subsection{The dust distribution}
\label{dust_distribution}

In radiative transfer models of spiral galaxies, dust is usually assumed to
be distributed in a smooth exponential disk similar to that adopted for stars.
At the reference wavelength of the V-band, the extinction coefficient of the 
smooth disk can be written as
\begin{equation}
k_\mathrm{V}(r,z)=\frac{\tau^\mathrm{f. o.}_\mathrm{V}}{2 \; z_\mathrm{d}} 
  \exp\left[-\frac{r}{h_\mathrm{d}} -\frac{|z|}{z_\mathrm{d}}\right],
\label{dustdisk}
\end{equation}
where $\tau^\mathrm{f. o.}_\mathrm{V}$ is the central, face-on, optical depth of the 
dust disk, and $h_\mathrm{d}$ and $z_\mathrm{d}$ are the radial and 
vertical scalelengths. 

If the dust properties are constant within the disk, Eq.~\ref{dustdisk}
can be integrated over the volume to give a total dust mass
\begin{equation}
M_\mathrm{dust}=\int k_\mathrm{V} dV \left(\frac{G/D}{m_\mathrm{H}}
\frac{\tau_\mathrm{V}}{N_\mathrm{H}}\right)^{-1} =
2\pi h_\mathrm{d}^2 \tau^\mathrm{f. o.}_\mathrm{V} 
\left(\frac{G/D}{m_\mathrm{H}}
\frac{\tau_\mathrm{V}}{N_\mathrm{H}}\right)^{-1},
\label{dmass}
\end{equation}
where $m_\mathrm{H}$ the mass of a hydrogen atom, $G/D$ is the gas (H) to 
dust mass ratio and $\tau_\mathrm{V}/N_\mathrm{H}$ is the V-band optical 
depth per unit H column density. The last two quantities are provided by
the adopted dust model.  As in \citet{BianchiSub1999}, I use 
$\tau^\mathrm{f. o.}_\mathrm{V}$ as an indicator of the amount of dust in 
a model, regardless if part of the dust is distributed in a clouds.
That is, a dust distribution labelled by a given value of 
$\tau^\mathrm{f. o.}_\mathrm{V}$ has the same $M_\mathrm{dust}$ as a 
smooth model with that central face-on optical depth.

A fraction $f_\mathrm{c}$ of the dust (gas) mass is assumed to be distributed 
in clumps, while the rest is in the smooth disk. Thus, in the presence of a
clumpy distribution, the smooth disk has a central face-on optical depth 
$(1-f_\mathrm{c}) \tau^\mathrm{f. o.}_\mathrm{V}$. In TRADING, the clumpy distribution 
consists of a collection of spherical, homogeneous clouds. Following
\citet{BianchiSub1999}, their properties are modelled on those of Giant 
Molecular Clouds (GMCs).

The mass distribution of GMC in the Milky Way and in a few other
Local Group galaxies is well described by a single power law,
\begin{equation}
\frac{dN}{dM} \propto M^{1-\gamma},
\label{mass_spectrum}
\end{equation}
over a wide range of cloud masses \citep{BlitzProc1999,BlitzProc2007}. 
For $\gamma < 2$, the H mass of a cloud $i$ can be derived with the MC
method from
\[
M_i=\left(M_\mathrm{inf}^{2-\gamma}+\mathcal{R} (M_\mathrm{sup}^{2-\gamma}-
M_\mathrm{inf}^{2-\gamma})\right)^\frac{1}{2-\gamma},
\]
with $M_\mathrm{inf}$ and $M_\mathrm{sup}$ the minimum and maximum
mass allowed in the spectrum.  The total number of clouds $N_\mathrm{c}$ 
is determined by requiring the conservation of the dust mass in the
clumpy component, i.e.
\[
(G/D)^{-1} \sum_i^{N_\mathrm{c}} M_i \approx f_\mathrm{c} M_\mathrm{dust}
\]
(due to the random nature of the cloud mass derivation, exact equality
is not assured).

Observations also suggest that the gas surface density of the clouds
is independent of the cloud mass, with a mean value $\Sigma$ that changes 
by less than a factor of two from galaxy to galaxy \citep{BlitzProc2007,
RosolowskyApJ2003}. If a constant $\Sigma$ is assumed, the 
radius $R$ follows directly from the cloud mass.
A constant $\Sigma$ also implies a constant optical depth along a cloud
diameter. For a homogeneous spherical cloud, the V-band extinction 
coefficient is then
\begin{equation}
k_\mathrm{V} = \frac{3}{4} \frac{\Sigma}{m_\mathrm{H} R}
\frac{\tau_\mathrm{V}}{N_\mathrm{H}}.
\label{kcloud}
\end{equation}

As for the clouds spatial distribution, it is assumed that it 
follows a double exponential as in Eq.~\ref{dustdisk}. The location
of each cloud is also used to define the location of the spherical
sources of radiation described in Sect.~\ref{stellar_emission}.

\subsection{The dust adaptive grid and the random walk}
\label{grid}

The analytical description of the dust distribution outlined so far 
is used to fill up the dust adaptive grid. A 3-D binary tree 
(an {\em octree}) is constructed subdividing recursively each volume 
element (a cell) into eight {\em children}. First, the dust distribution 
is enclosed in a cubic volume of size $D$ (the {\em root} cell). Then, 
the condition
\begin{equation}
\int_\mathrm{cell} k_\mathrm{V} dV \le E 
\int_\mathrm{D^3} k_\mathrm{V} dV,
\label{splitting}
\end{equation}
is checked, with $E<1$ a grid refinement parameter 
\citep{WolfA&A1999,KurosawaA&A2001}. I have used 
$E=10^{-5.5}$ for the simulations presented in this paper. If the condition
is not met, the cell is split into eight cubes and the condition checked
again recursively until it is met. If the condition is met, subdivision
is ended and a {\em leaf} cell is found. If $L$ subdivisions have taken
place, the {\em leaf} cell has size $D/(2^{L})$~\footnote{As an example, 
let's consider
a homogeneous {\em root} cube: setting, e.g., $E=8^{-3}$ is equivalent 
to produce a regular grid of $8^3$ {\em leaf} cells (descending down
to level $L=3$), each of size $D/8$. Using a cubic {\em root} volume to 
enclose a thin disk-like structure may seem inadequate, since regions 
distant from the galactic plane are empty. However, these regions are
enclosed by a limited number of low-level {\em leaf} cells. Furthermore, 
the cubic geometry makes the structure of more general application and is 
more effective in the description of the spherical clouds in the clumpy
distribution than cells with a reduced size along $z$.}.
For each cell, information on its boundaries are stored, together with
pointers to its {\em father} cell (unless it is the {\em root} cell) and 
to its eight {\em children} (unless it is a {\em leaf} cell). The local 
value of $k_\mathrm{V}$ is assigned to each {\em leaf} cell. 

\begin{figure}
\resizebox{\hsize}{!}{ \includegraphics{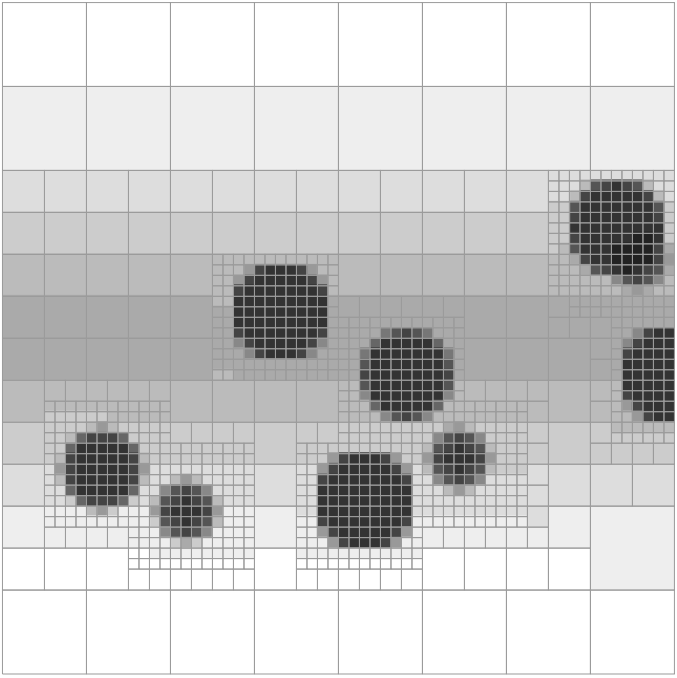} }
\caption{A cut through the binary dust grid used in Sect.~\ref{clumping}, 
perpendicular to the galactic disk. Since mean values are stored
in each cell, the density on the border of clouds does not drop abruptly to
the value of the smooth medium. The size of the box is 1.5 kpc x 1.5 kpc.
}
\label{clumps}
\end{figure}

The criterion in Eq.~\ref{splitting} is based on the mean opacity 
of a cell (equivalently, on the density, if the dust properties are
identical throughout space, as in this work). Since it is not tied 
to the local radiation field (which is not known before the radiative 
transfer calculations), it may fail in producing an adequate
cell resolution in the presence of very strong field gradients.
A high resolution is needed for the case of clouds with internal 
point sources, and could in principle be achieved by using a smaller 
value for $E$. However, this would result in a larger number of cells 
and in an unnecessary finer sampling of the diffuse medium\footnote{For 
smooth models, and for the smooth medium in the clumpy models presented 
here, the chosen $E$ value ensures enough spatial resolution for the 
convergence of the results.}.
To avoid this, the grid algorithm allows to force splitting in cells:
if the cloud diameter is smaller than the dimension of a cell of level
$L$, the {\em leaf} cells can be required to split further by $l$ sublevels,
down to level $L+l$, 
thus providing at least $2^{l-1}$ resolution elements along a cloud diameter.
An example of the binary structure of the dust grid is shown in 
Fig.~\ref{clumps}.

The optical depth of the dust distribution along any given direction is 
found from the intercepts of the {\em photon} path with the boundaries of the 
{\em leaf} cells in the grid \citep[for more details, see][]{KurosawaA&A2001}. 
The {\em leaf} cell that contains the position of emission is found by 
recursively ascending the binary three from the root: the point location 
algorithm of \citet{FriskenGraphics2002} is used. The coordinates of the 
intercept of the {\em photon} path with the boundaries of this first cell are 
used to locate the next cell along the path. This neighbour cell could be 
found again by ascending the tree from the root. An alternative scheme
would consist in finding the smallest non-{\em leaf} cell that is a common 
ancestor for both a cell and its neighbour: the tree is then descended
from the first cell to the common ancestor, and then ascended again from the
ancestor to the cell neighbor. This second implementation is more efficient
\citep{FriskenGraphics2002}. If $dl$ is the length of the intercept,
the optical depth along the path in the cell is
\[
d\tau_\lambda=A_\lambda k_\mathrm{V} dl,
\]
where $\lambda$ is the wavelength of the {\em photon} and $A_\lambda$
is the dust extinction law from the adopted model. The process is then 
repeated and the optical depth along the path $\tau_\lambda$ accumulated,
until the {\em photon} suffers the first scattering event (or the boundary of
the root cell is reached).

The location of the scattering event is determined randomly, by finding
the optical depth 
\begin{equation}
\tau = -\log (1-\mathcal{R}),
\label{tau}
\end{equation}
and stopping the path in the grid where $\tau_\lambda=\tau$. If 
$\tau$ is larger than the total optical depth along the chosen direction,
the {\em photon} escapes dust. To ensure that all {\em photons} contribute 
to the scattered flux, even in the case of low global opacity, scattering
may be {\em forced} \citep{CashwellBook1959}. Typically, the
first scattering event is forced, while the following are determined
according to Eq.~\ref{tau} \citep{BianchiApJ1996,GordonApJ2001}. The 
same scheme is adopted here.

In the scattering position, the {\em photon} propagates along a new 
direction, which is derived randomly from the adopted scattering
phase function. Here I use the \citet{HenyeyApJ1941} phase 
function, dependent on the scattering asymmetry parameter 
$g_\lambda=\langle\cos(\theta)\rangle$, with $\theta$ the angular deviation
between the new and the old directions.  The \citeauthor{HenyeyApJ1941} 
phase function is a reasonably good approximation for the wavelengths
considered here \citep{DraineApJ2003}. The values for $\omega_\lambda$ 
and $g_\lambda$ are provided by the adopted dust model. 

The scattered {\em photon} has its {\em weight} reduced by the albedo
$\omega_\lambda$, while the fraction $(1-\omega_\lambda)$ is absorbed.
The absorbed energy could be stored in the {\em leaf} cell where scattering
takes place and used later to derive dust emission
\citep{BianchiA&A2000b,GordonApJ2001}. However, it is more efficient to
store the absorbed energy along all the cells crossed by
a {\em photon} rather than at the end point only, a concept introduced by 
\citet{LucyA&A1999} and described in \citet{NiccoliniA&A2003} for both the 
non-forced and forced scattering cases. Using this formalism, TRADING 
accumulates the amount of energy absorbed as a function of the wavelength,
$W_\lambda$, from which the local ISRF is derived,
\[
J_\lambda=\frac{1}{4 \pi k_\lambda (1-\omega_\lambda)}\;
\frac{W_\lambda}{\Delta V},
\] 
with $\Delta V$ the cell volume. For each {\em leaf} cell, the average of 
$J_\lambda$ is stored in $N_\mathrm{SED}^\mathrm{stars}$ contiguous 
wavelength bins, logarithmically spaced in the range 
$[\lambda_\mathrm{min}^\mathrm{stars},\lambda_\mathrm{max}^\mathrm{stars}]$. 

After the first scattering, the whole procedure is reiterated, until
the {\em photon weight} falls below a limit value ($10^{-4}$ is used here).
A number $N_\mathrm{phot}$ of {\em photons} are run to provide high 
signal-to-noise images and $J_\lambda$ values in the grid. The model can 
be normalised to physical values by multiplying the outputs by 
$L_\mathrm{bol}/N_\mathrm{phot}$, $L_\mathrm{bol}$ being the bolometric 
luminosity assumed for the stellar sources.

Simulated images of dust-extinguished starlight can be also produced,
for all the $N_\mathrm{SED}^\mathrm{stars}$ bins or at specific
values of $\lambda$. Any viewing direction is possible. This is
achieved using the {\em peeling-off} technique of \citet{YusefZadehApJ1984}, 
in which each emission and scattering event contributes to the surface 
brightness distribution. The technique allows to produce high 
signal-to-noise results without the need of collecting {\em photons} in
a broad angle band \citep[see, e.g.,][]{BianchiApJ1996}, thus preventing
image smearing when there is a strong gradient with the viewing direction 
(as for, e.g., a disk seen edge-on).

Once that the radiation field has been determined, it may result that 
some cells contribute negligibly to the total absorbed (and emitted) 
energy. To avoid unecessary calculation, the grid is reduced in size
by adopting a criterion similar to that in \citet{MisseltApJ2001}:
cells are ordered according to the amount of energy absorbed within
them, and a threshold is derived so that all cells below the threshold
contributes to a fraction $f_\mathrm{abs}$ of the total absorbed energy. 
The grid is recursively accessed to find {\em fathers} of {\em leaf} cells
which have all eight {\em children} below the threshold. When such cells 
are found, the {\em children} are cut and their amount of absorbed energy 
is transferred to the {\em father}, which becomes {\em leaf}. Instead, 
when {\em leaf} cells below the threshold are isolated, they are excluded 
from the calculation of dust emission. When a clumpy dust distribution is 
considered, even a small value of $f_\mathrm{abs}$ may result in an 
appreciable reduction of the memory occupation of the grid and of
execution time of the emission code. In this work, I have used 
$f_\mathrm{abs}=0.1\%$. 
No significant change in the results is produced by the adoption of
this threshold.

\subsection{Dust emission}
\label{dustemi}

When considering the transfer of radiation through absorption and
scattering, it is not necessary to know the details on the dust
composition and size distribution, but it is sufficient to consider
mean properties \citep{WolfApJ2003}: these are the values $A_\lambda$, 
$\omega_\lambda$ and $g_\lambda$ used in the previous section.

Dust emission, instead, depends on the properties of each grain. For
an ensemble of spherical grains of $N_\mathrm{mat}$ different
materials, each characterised by a distribution $n_k(a)$ for
the radius $a$, the emission coefficient (energy emitted per unit 
volume, time, wavelength and solid angle) for thermal equilibrium 
emission can be written as
\begin{equation}
j_\lambda =  k_\mathrm{V}
\;\;
\frac{
\displaystyle{
\sum_{k=1}^{N_\mathrm{mat}} \int n_k(a)\; a^2\; 
Q^\mathrm{abs}_k(a,\lambda)\; 
B_\lambda(T_\mathrm{d}) \;da
}
}{
\displaystyle{
\sum_{k=1}^{N_\mathrm{mat}} \int n_k(a)\; a^2\; 
Q^\mathrm{ext}_k(a,\mathrm{V})\; da,
}
}
\label{ecoeff}
\end{equation}
with $Q^\mathrm{ext}_k$ and $Q^\mathrm{abs}_k$ the extinction and absorption
(emission) efficiencies (i.e.\ the cross-sections divided by $\pi a^2$), 
$B_\lambda$ the Planck function and  $T_\mathrm{d}$ the equilibrium 
temperature of a grain, dependent on the radius $a$ and on the material $k$.
In TRADING, the size distribution for each material $k$ is sampled with 
$N_a^k$ contiguous bins logarithmically spaced between the minimum and
maximum radii of the adopted grain distribution. Equation~\ref{ecoeff} thus 
becomes
\begin{equation}
j_\lambda = k_\mathrm{V}
\sum_{k=1}^{N_\mathrm{mat}} \sum_{j=1}^{N_a^k} Q_{jk}(\lambda)
\;\mathcal{B}_{jk}, 
\quad\quad\quad
\mathcal{B}_{jk}= B_\lambda(T_\mathrm{d}(a_j,k)),
\label{jlcomp}
\end{equation}
where $Q_{jk}$ is the absorption cross section integrated over the
$j$-th radius bin of width $\Delta a_j$, normalised by the V-band
extinction cross section\footnote{A
table with $Q_{jk}(\lambda)$ is computed from the adopted dust model
and provided to the radiative transfer code. Ancillary tables are
produced by averaging $Q_{jk}$ over the wavelength bins used for
the sampling of $j_\lambda$ (Eq.~\ref{jlcomp}), $J_\lambda$ 
(Eq.~\ref{eqtempe}) and $J_\lambda^\mathrm{s. a.}$.
},
\begin{equation}
Q_{jk}=\frac{
\displaystyle{
n_k(a_j)\; a_j^2\; Q_k^\mathrm{abs}(a_j,\lambda)\; \Delta a_j
}
}
{ 
\displaystyle{
\sum_{k=1}^{N_\mathrm{mat}} \int n_k(a)\; a^2\; 
Q_k^\mathrm{ext}(a,\mathrm{V})\; da}.
}
\label{dusttable}
\end{equation}

At thermal equilibrium, $T_\mathrm{d}$ can be found equating the radiation 
absorbed from the ISRF $J_\lambda$ to the dust 
emission. For the $j$-th grain of material $k$, the thermal balance can 
be written as
\begin{equation}
\sum_{i=1}^{N^\mathrm{stars}_\mathrm{SED}} J_\lambda Q_{jk}(\lambda) 
\Delta\lambda_i =
\int B_\lambda(T_\mathrm{d}(a_j,k)) Q_{jk}(\lambda) d\lambda,
\label{eqtempe}
\end{equation}
where $\Delta\lambda_i$ is the width of the $i$-th bin used to store 
$J_\lambda$. To save computing time, the integral in Eq.~\ref{eqtempe} 
is tabulated, and the temperature $T_\mathrm{d}(a_j,k)$ is found by 
interpolation.

For small grains ($a\la 0.01 \mu$m) heating is stochastic and it
is not possible to define an equilibrium temperature. Instead, a 
given grain is characterised by a temperature distribution 
$P(T)$, dependent on the grain characteristics and
on the heating field. Following the method of \citet{GuhathakurtaApJ1989},
$P(T)$ can be derived from the transition matrix $A_{f,i}$,
that gives the probability per unit time that a dust grain is heated
(or cooled) from the enthalpy state $H_i$ to the enthalpy state $H_f$.
In the notation used so far, for a given grain, the matrix elements due 
to heating by the ISRF  are ($f>i$) 
\[
A_{f,i} \propto J_\lambda \frac{Q_{jk}(\lambda) \; \Delta H_f}{(H_f-H_i)^3},
\]
where $\Delta H_f$ is the width of the $f$ enthalpy state, 
$\lambda=hc/(H_f-H_i)$ is the wavelength associated to the transition,
and the value of $J_\lambda$ is obtained by interpolating the ISRF provided
by the radiative transfer code
\citep[see][for a full description of the matrix 
elements]{MisseltApJ2001}. Considering only cooling terms from level $f$+1 
to level $f$ \citep[the {\em thermal continuous} cooling approximation;][]
{DraineApJ2001}, it is
\[
A_{f,f+1}\propto\frac{
\int Q_{jk} (\lambda) \; B_\lambda(T_{f+1}) d\lambda,
}
{H_{f+1}-H_f}
\]
where $T_{f+1}$ is the temperature associated to state $f$+1.
In TRADING, a temperatures range for $P(T_\mathrm{d})$ is selected, and
$N_T$ states are defined, logarithmically spaced in temperature. The 
enthalpy associated to each state is derived by integrating over the
temperature the specific heat adopted for the grain materials. The same
temperature range and $N_T$ are used for all grain, regardless of the
local ISRF intensity and spectrum. This allows to compute tables for 
the cooling terms and for the multiplicator of $J_\lambda$ in the heating 
terms beforehand, thus saving computing time. For a grain of material
$k$ and radius $a_j$ that undergoes stochastic heating, the $\mathcal{B}_{jk}$
term in Eq.~\ref{jlcomp} becomes
\begin{equation}
\mathcal{B}_{jk}= \sum_{t=1}^{N_T} P_t B_\lambda(T_{t}).
\end{equation}

In each cell, the emission coefficient $j_\lambda$ is computed by summing 
the contribution of all sizes and materials in the dust distribution.
The calculation is carried out for $N_\mathrm{SED}^\mathrm{dust}$ 
contiguous wavelength bins, logarithmically spaced in the range
$[\lambda_\mathrm{min}^\mathrm{dust},\lambda_\mathrm{max}^\mathrm{dust}]$.

Subsequent calculations are carried out using a no-scattering MC 
radiative transfer procedure. The MC allows to take into account 
self-absorption, i.e. the possibility that dust-emitted radiation
is again absorbed by dust. This could be relevant in some astrophysical
situations, when MIR radiation travels through a very dense
dusty medium. 

In principle, the cell where a {\em photon} (of unit {\em weight}) is 
emitted should be selected randomly from the amount of energy emitted 
within it, and the wavelength determined randomly from the local 
$j_\lambda$ spectrum.
However, it is not straightforward (or efficient in terms of computing 
time) to access the binary tree in this way. Furthermore, the procedure 
would also require to store $j_\lambda$ in the memory for each cell. 
Also, since 
scattering is not included, calculations can be carried out at once for
all wavelengths. Thus in TRADING, once $j_\lambda$ is derived, a few 
{\em polychromatic} {\em photons} are emitted (five are sufficient for
the simulations presented in this paper), with a $\lambda$-dependent 
{\em weight} proportional to $j_\lambda\Delta V$ ($\Delta V$ 
is the cell volume). The emission coordinates within a cubic (homogeneous) 
{\em leaf} cell and the direction propagation are determined in the usual 
way. As the
{\em photon} traverses cells, a local ISRF $J_\lambda^\mathrm{s. a.}$
can be computed from self-absorption in the wavelength range 
$[\lambda_\mathrm{min}^\mathrm{dust},\lambda_\mathrm{max}^\mathrm{dust}]$,
analogously to what done for stellar {\em photons}. In case, the 
wavelength range could be limited by a maximum wavelength
$\lambda_\mathrm{max}^\mathrm{s. a.}$, if self-absorption is insignificant
for $\lambda>\lambda_\mathrm{max}^\mathrm{s. a.}$, and a different
number of bins over which $J_\lambda^\mathrm{s. a.}$ is stored,
$N_\mathrm{SED}^\mathrm{s. a.}$, can be selected. Selecting a restricted
wavelength range for self-absorption may become necessary when computer 
memory is a concern, since, contrary to $j_\lambda$, it is necessary to
store $J_\lambda^\mathrm{s. a.}$ for all cells before carrying on the
calculations. 

\begin{figure*}
\centering
\includegraphics[height=8.5cm]{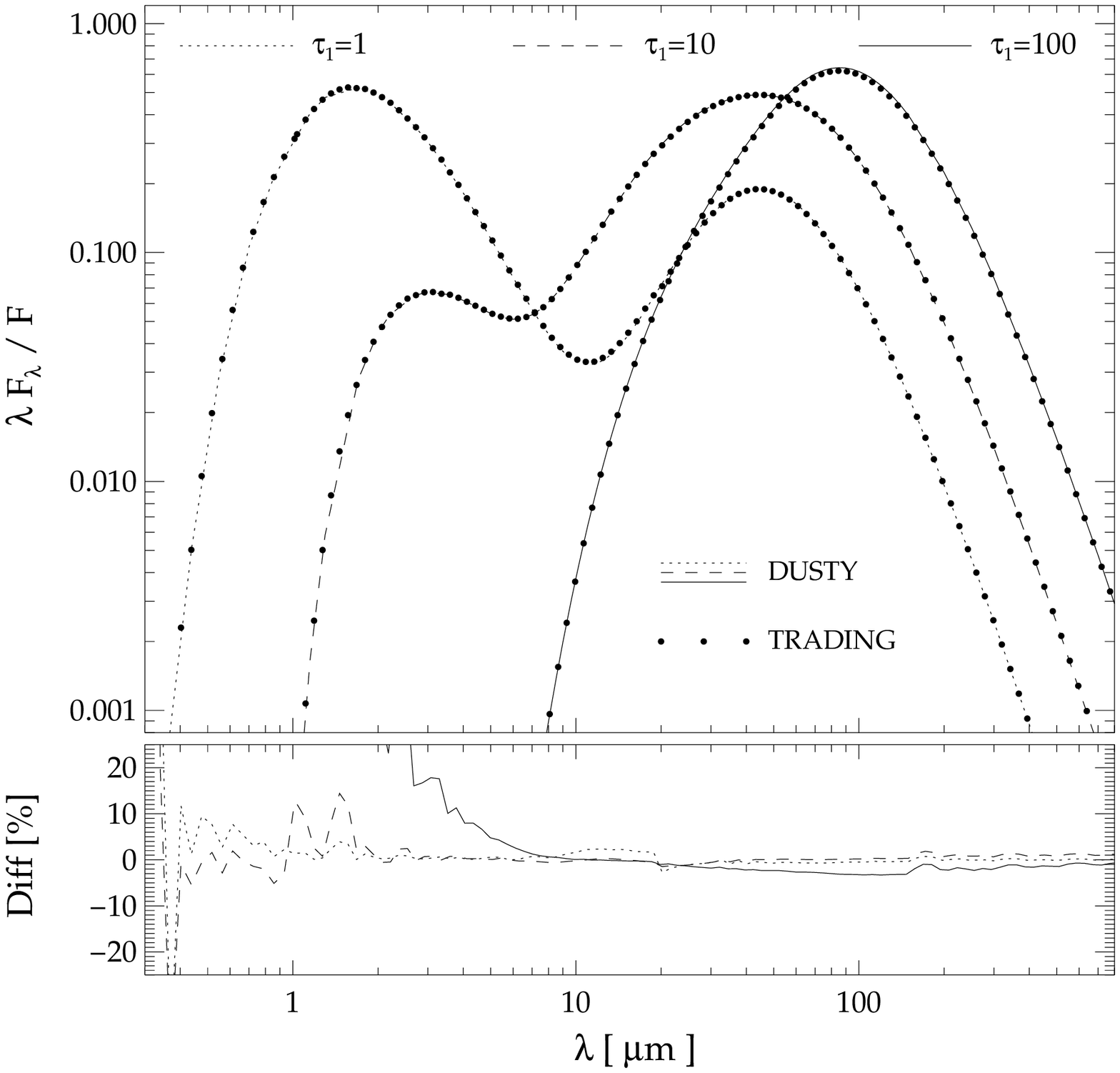}
\includegraphics[height=8.5cm]{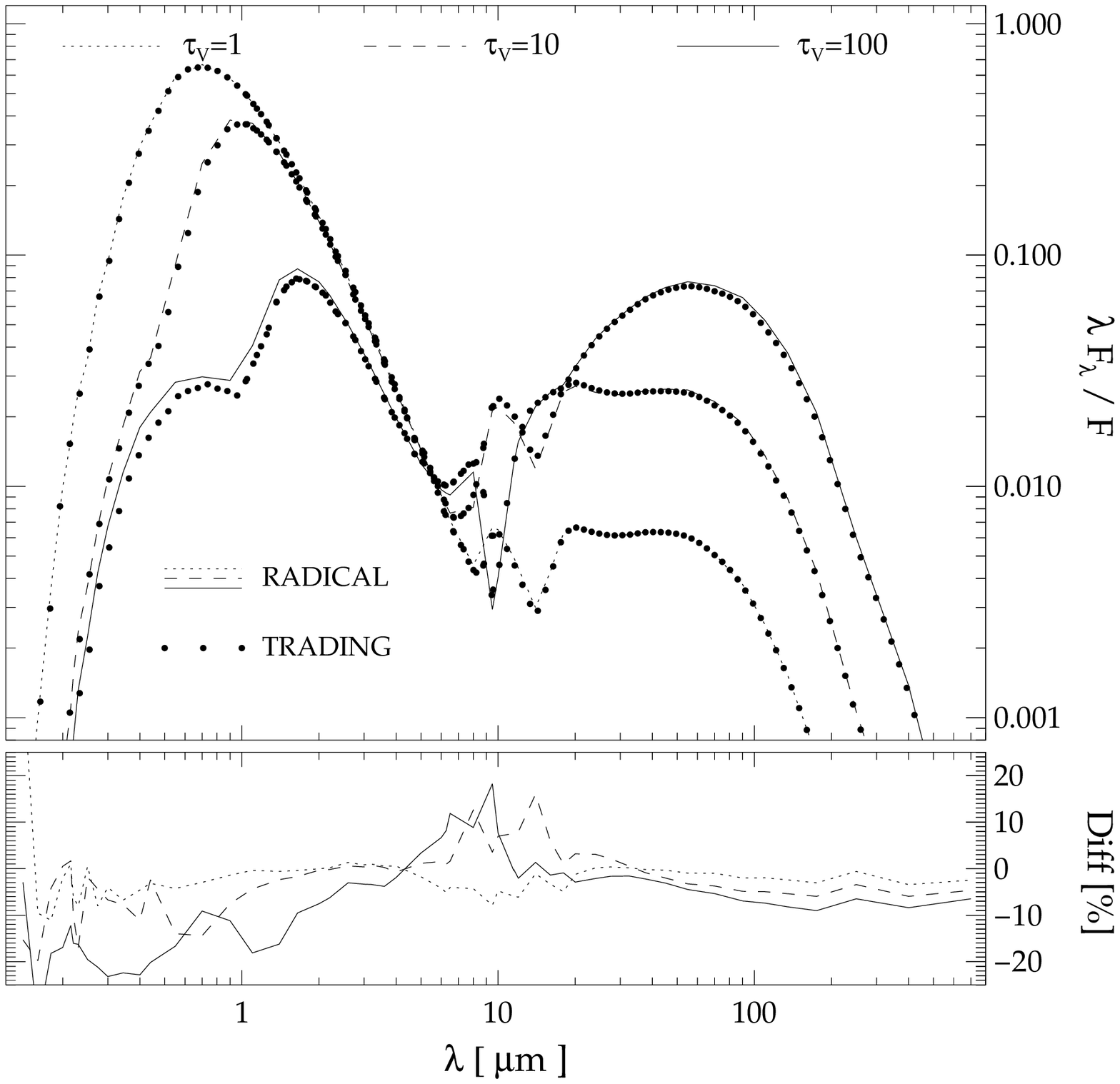}
\caption{Tests of the TRADING code. On the left, comparison with the
DUSTY solutions of the benchmark problems of \citet{IvezicMNRAS1997a} 
for a homogeneous sphere. On the right, comparison with the RADICAL
solutions for the benchmark problems of \citet{PascucciA&A2004}. In 
all cases, the SED has been normalised to the total luminosity;
the relative difference has been computed with respect to the benchmark
solutions. See text for details.}
\label{benchsed}
\end{figure*}

Once  $J_\lambda^\mathrm{s. a.}$ is derived, its contribution to dust 
heating is added to that of $J_\lambda$, and a new emission coefficient
$j_\lambda$ can be computed. 
The entire process is repeated until the total output of dust emission 
(i.e. the dusty SED integrated over the solid angle)
does not change by more than a given fraction, typically a few percents. 
The convergence check, as well as the splitting of TRADING in the
extinction and emission parts, could be avoided by using the
{\em immediate re-emission} concept of \citet{BjorkmanApJ2001}.
With this procedure, the absorption of a stellar {\em photon} is immediately 
followed by the emission of an infrared {\em photon} from dust, with 
the wavelength sampled from a distribution that corrects itself at
each absorption event; eventually the distribution of infrared
{\em photons} converges to the correct distribution, without the need for 
a separate calculation of the absorption and emission processes. Although 
the method has a solid mathematical ground in the case of thermal 
equilibrium heating and in MC procedures where {\em photons} have {\em weights},
it is not clear yet if it can be applied with the \citet{LucyA&A1999}
estimator for $J_\lambda$ or in case of stochastic heating
(\citealt{BaesNewA2005}; but see \citealt{JonssonMNRAS2006} 
for a different view on the second issue).
Because of this, I have preferred the current scheme.

As for starlight, images of dust-emitted radiation can be obtained from any 
viewing direction, for all the $N_\mathrm{SED}^\mathrm{dust}$ bins
or at specific values of $\lambda$, using the {\em peeling-off} technique
from each emission point in the MC procedure.

\subsection{Benchmarking}
\label{bench}

The code results have been tested on a few radiative transfer solutions: 
the outputs from the MC radiative transfer code of \citet{BianchiApJ1996}; 
the description of the radiation field due to a point source in the center 
of a finite scattering sphere \citep{SiewertJQSRT1979}; the benchmark 
problems in 1D geometry of \citet{IvezicMNRAS1997b}; the benchmark
problems in 2D geometry of \citet{PascucciA&A2004}. 
I describe here the comparison with the last two benchmark cases.

\citet{IvezicMNRAS1997b} describe a set of radiative transfer problems for 
a spherical dust distribution heated by a central point source. I have
compared the output of TRADING with the case of a homogeneous sphere. 
The solutions have been obtained in numerical format using the publicly 
available radiative transfer code 
DUSTY\footnote{\tt http://www.pa.uky.edu/\~{}moshe/dusty/}, based on the
method of \citet{IvezicMNRAS1997a}. DUSTY is one of the radiative
transfer codes used in \citet{IvezicMNRAS1997b}.

A grid for a homogeneous sphere of radius $R_\mathrm{out}$ was first 
constructed. The dust distribution of the benchmark cases has a inner 
cavity for $R_\mathrm{in}=R_\mathrm{out}/1000$. Most of the sphere 
volume is sampled with cells of level $L=7$ (thus, having a size of 
$2 \times R_\mathrm{out}/128$), but finer subdivisions (down to $L=14$)
are used for a proper rendering of the inner and outer boundaries. 
In total, the grid consists of $6\times10^5$ {\em leaf} cells.
The extinction coefficient (at the reference wavelength of 1 $\mu$m) is 
given by the optical depth of the model along the radius, 
$k_1=\tau_1/(R_\mathrm{out}-R_\mathrm{in})$.

DUSTY only deals with a single grain of mean properties and the scattering
is assumed to be isotropic. Thus, $g_\lambda=0$ has been set in TRADING,
and $A_\lambda$ and $\omega_\lambda$ have been chosen as in 
\citet{IvezicMNRAS1997a}.

In the benchmark cases, radiation comes from a central point source, 
emitting as a blackbody at $T_\star=2500$~K. That ISRF has been sampled 
with 50 bins in the range $[0.3,20]$ $\mu$m. DUSTY solutions are 
scale-free \citep{IvezicMNRAS1997a}, and it is not necessary to provide 
the absolute luminosity. Instead, for the comparison with TRADING, a 
physical value for $R_\mathrm{out}$ was assumed, and the bolometric 
luminosity $L_\mathrm{bol}$ was derived with the relations provided in 
\citet{IvezicMNRAS1997a} and \citet{IvezicMNRAS1997b}. 
Dust radiation has been sampled with 100 bins in the range $[1,1000]$ $\mu$m.
The same sampling has been used for the contribution of self-absorption to 
the ISRF.
A total of $10^6$ photons have been used.

The result from the comparison is shown in the left panel of 
Fig.~\ref{benchsed} for the cases $\tau_1=1,10$ and $100$. As TRADING 
deals with extinction and emission separately, the stellar and dust 
SEDs have been summed together in the overlapping wavelength range. 
The inclusion of self-absorption has been necessary for a correct 
solution of the two high optical depth cases: 
convergence to 0.5\% was required. 
The agreement with the benchmark cases is satisfactory, with TRADING
differing from DUSTY by only a few percents in most cases. 
The agreement is worse at shorter wavelengths, for several reasons:
the intrinsic noise in the MC procedure; the difficulty of properly 
sampling the integral of a steeply rising spectrum (Eq.~\ref{mcspectrum})
with tables; 
the lack of ad-hoc optimisations for high optical depths 
\citep[see, e.g., ][]{JuvelaA&A2005}.
In particular, TRADING diverges significantly from the DUSTY solution 
for the $\tau_1=100$ case, but only for $\lambda\la 3\mu$m and very 
low flux levels ($\la 10^{-7}$ in the units of Fig.~\ref{benchsed}).
Calculations for these low fluxes in the higher optical depth case also
pose problems for DUSTY: numerical errors had to be prevented
by requiring a significantly higher numerical accuracy than in the
other cases.
Nevertheless, the radiation field appears to be computed correcly by
TRADING in all cases, as shown by the bulk of dust emission in the SED.
Also, the surface brightness profiles of stellar and dust radiation 
show a similar agreement. 

The geometry analysed so far is equivalent to what will be used 
in Sect.~\ref{clumping} for embedded stellar sources at the center of 
GMCs. Unfortunately, in simulations of galactic disks it is not possible 
to have, for each cloud, the same level of spatial resolution as in the 
comparison presented here. Thus, it will become necessary to check
for the convergence of the simulation outputs.

A benchmark problem with a more compelling geometry is provided by 
\citet{PascucciA&A2004}. They use an axially symmetric dust disk,
extending from an inner cavity of 1 AU to an external boundary of
1000 AU, with a steeply rising density gradient close to the center.
The dust grid I have used for this problem has been obtained
with the procedure described in Sect.~\ref{grid}, but forcing cell
splitting in the inner 50 AU down to $L=14$ close to the center.
The grid consists of $1.1\times10^6$ {\em leaf} cells. Dust properties 
are provided by the authors, and isotropic scattering is assumed.

As for the previous case, the heating source is point like and located 
in the center, with $L_\mathrm{bol}=1 L_{\sun}$ and blackbody emission
at $T_\star=5800$~K. For this case, the ISRF has been sampled 
with 50 bins in the range $[0.12,10]$ $\mu$m. All the other
parameters are the same as for the comparison with DUSTY.

As a reference, I have used the solution of the benchmark problems
provided by RADICAL \citep{DullemondA&A2000}, one of the codes used 
in \citet{PascucciA&A2004}. The result of the comparison with their
{\em edge-on} case (inclination $i=77.5^\circ$ from the polar axis) 
is shown in the right panel of Fig.~\ref{benchsed}. Again,
the agreement is satisfactory. For most wavelengths the TRADING
solution is within 10\% of the RADICAL solution, a difference
only sligthly larger than what achieved by comparing the outputs
of the various codes used in \citet{PascucciA&A2004}. Only the
high optical depth model shows a larger discrepancy on the stellar
side of the SED. This is due to the limited resolution of the grid.
Better agreement is obtained for the more face-on models.

As pointed out by \citet{JonssonMNRAS2006}, regular or adaptive 3D
orthogonal grids are not ideally suited for a comparison with the 
benchmark solutions described here. In fact, no benchmark solution 
is available yet for the case of diffuse sources cospatial with dust,
a configuration more suitable for the study of internal extinction 
and emission in galaxies. This is the subject of a recently started
project on comparing different codes that cater to the study of 
radiative transfer in galaxies.

\section{The dust model}
\label{dust}

The dust grain model of \citet{DraineApJ2007b} has been adopted for the 
simulations presented in this work. The model consists in a mixture of
silicate and carbonaceous grains. The size distributions was derived
by fitting the extinction law and infrared emission from dust as observed 
in our Galaxy \citep[][the model for the $R_\mathrm{V}=3.1$ extinction
law is used here]{WeingartnerApJ2001a,LiApJ2001,DraineARA&A2003}. 

The dust grain extinction and absorption cross-sections have been derived 
using the Mie theory for spherical particles (\citealt{MieAnnPhys1908}; see 
also \citealt{BohrenBook1983}). The optical properties of amorphous 
{\em astronomical} silicates are used for the silicate component and
those of graphite for the larger carbonaceous grains.

For small carbonaceous grains, the absorption cross sections are
derived from the properties of PAH materials, the most probable 
responsible of the MIR emission features \citep{LiApJ2001,DraineApJ2007b}.
In the model used here, these grains contain 4.6\% of the mass of the
carbonaceous components. The properties of PAH grains also depend on
the ionization state, and the ionization fraction of \citet{DraineApJ2007b} 
is used. The absorption cross section of the carbonaceous components is 
assumed to change smoothly from that of PAHs to that of graphite 
at a radius of 0.005 $\mu$m: the smooth transition provides
a continuum absorption component to the PAH features for small grains,
while for $a\ga0.01$ $\mu$m the graphite component becomes dominant
\citep{DraineApJ2007b}.

With the above recipe, I have computed the mean extinction law
$A_\lambda$, albedo $\omega_\lambda$ and asymmetry parameter
$g_\lambda$ which are used for the radiative transfer of
stellar radiation ($A_\lambda$ is also used for dust radiation 
if self-absorption is included); and the elements of the $Q_{jk}$ 
table of Eq.~\ref{dusttable}, which are necessary to compute the
dust emission spectrum. In the table, $N_\mathrm{mat}=4$ grain
components are considered: silicate grains, with $N_a^{\mathrm{sil}}=18$
bins in radius from $a=3.5$\AA\ to $0.5 \mu$m; graphite grains,
with $N_a^{\mathrm{gra}}=10$ bins in radius from $a=0.01 \mu$m to 
$1 \mu$m; ionised carbonaceous grains (smoothly changing from
PAHs to graphite), with $N_a^{\mathrm{PAH}}=9$ bins in radius 
from $a=3.5$\AA\ to $0.01 \mu$m; neutral carbonaceous grains with 
the same bins as for ionised grains. The values of $N_a^{k}$ results 
in similar radial bins for both carbonaceous and silicate grains. 
The sampling of the different components was chosen by means of 
trial and was dictated by the 
necessity of having the smallest possible number of bins (since
the computing time depends on the sums in Eq.~\ref{jlcomp}) still
able to provide a reasonably accurate estimate of dust emission.

Computing time is an issue especially for grains that undergoes
stochastic heating. As noted by \citet{MisseltApJ2001}, this
is the real bottleneck of the computation. For the models presented 
here, the calculation
of the temperature probability distribution $P$ is limited to
grains with $a\le 0.01\mu$m (26 bins out of 46), while the rest 
are assumed to emit at thermal equilibrium. The enthalpy states
are sampled with $N_T=80$ temperature bins, with 2.7K$<T<$2000K. 
The \citet{DraineApJ2001} models for the specific heats of bulk 
silicate and graphite have been used. While this temperature
sampling is sufficient for the smaller grains, it does 
not provide correct solutions for the larger grains exposed to
stronger ISRFs, whose $P$ is close to a delta function. When this
happens, thermal equilibrium emission is used. A more accurate
calculation would require a different temperature sampling
tailored over the heating conditions in each cell and for each
grain, as in the iterative algorithm described by 
\citet{MisseltApJ2001}. However, this is unpractical for the
large grid considered here, as the method
prevents the usage of pre-computed tables for the terms in the 
transition matrix and would result in too long computation times. 

\begin{figure}
\resizebox{\hsize}{!}{ \includegraphics{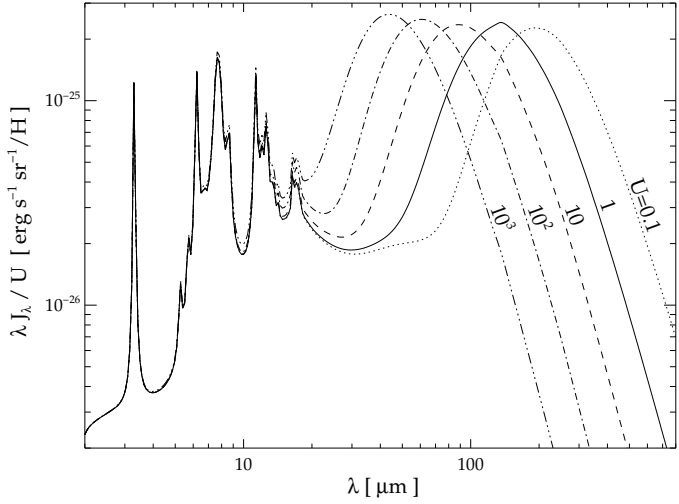} }
\caption{Dust emission spectra for the dust model of Sect.~\ref{dust},
normalised to the hydrogen column density. The heating spectrum is the
average Galactic ISRF, with intensity scaled by factor $U$.}
\label{isrf}
\end{figure}

Despite the approximations and the limited grain sampling, the method 
used in TRADING results in a reasonably good description of dust emission.
Fig.~\ref{isrf} shows the emission when heating is due to the average
Galactic ISRF (\citealt{MathisA&A1983}; see also 
\citealt{WeingartnerApJS2001}), 
scaled by different factors $U$. In the wavelength range where emission 
is due to stochastically heated grains, the agreement with the analogous 
models of \citet{DraineApJ2007b} is within $\approx10$\%. A similar
agreement with full models is achieved using the spectra of
Sect.~\ref{results}.

Finally, the dust model is characterised by $G/D = 90$ and 
$\tau_\mathrm{V}/N_\mathrm{H} = 4.9 \times 10^{-22}$~cm$^2$.

\section{An application: NGC891}
\label{results}

In this section, I present TRADING models for a galactic disk.
The physical quantities assumed in the simulations are tied
to the case of \object{NGC891}, an Sb-type edge-on galaxy whose SED
and surface brightness distributions have been subject of
extensive studies \citep{XilourisSub1998,PopescuA&A2000,
PopescuA&A2003a,AltonA&A2004,DasyraA&A2005}. In accordance with
these works, a distance of 9.5~Mpc is assumed.

\begin{figure}
\resizebox{\hsize}{!}{ \includegraphics{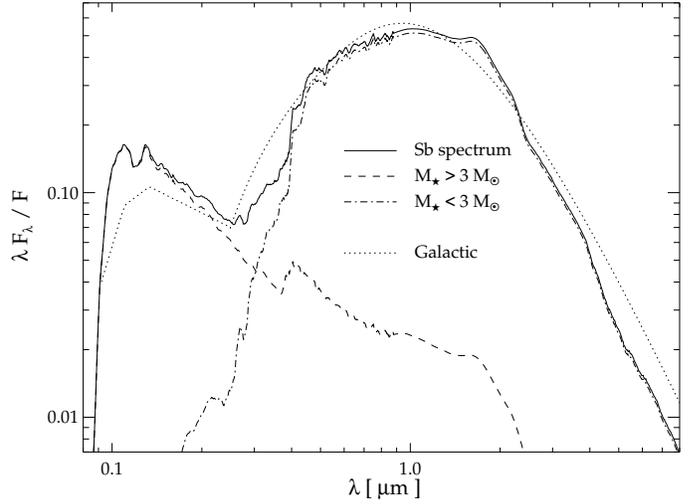} }
\caption{The stellar SED from the Sb template of \citet{FiocA&A1997},
normalised to the bolometric luminosity.
The total SED is shown together with the separate contribution of
stars of
low ($M_\star<3$~M$_{\sun}$) and high ($M_\star>3$~M$_{\sun}$) mass.
For the sake of presentation, the spectra have been smoothed over
the wavelength. The average Galactic ISRF is also shown for comparison.
}
\label{sb_sed}
\end{figure}

For stars, I have adopted a typical spectrum of an evolved galaxy.
It has been obtained by running the spectral synthesis code PEGASE.1 
of \citet{FiocA&A1997} with the parameters for their Sb template
\citep[see also ][]{LeithererPASP1996}, but without including a dust 
correction.  The intrinsic stellar spectrum is shown in Fig.~\ref{sb_sed}; 
it is similar in shape to the 
average Galactic ISRF. By setting an arbitrary cut at $3M_{\sun}$,
the spectrum has been decomposed into the separate contribution of low 
mass stars (emitting most of $\lambda \ga 0.3 \mu$m radiation, and
contributing to about $80\%$ of the bolometric luminosity), and
high mass stars (emitting most of UV radiation). This will allow to 
use separate spatial distributions for young OB stars and for the older 
population.  

In all the models, the stellar $J_\lambda$ is sampled with 
$N_\mathrm{SED}^\mathrm{star}$ = 20 bins in the range $[0.09,8] \mu$m.
Instead, dust emission is sampled with $N_\mathrm{SED}^\mathrm{dust}$ = 
150 bins in the range $[2,1500] \mu$m. For all the wavelength bins, images 
are produced for selected inclinations, and a global SED
is obtained by integrating over the maps. 

Models are compared to the SED of NGC891, which is known 
over a broad wavelength range: total fluxes are available in the UV 
\citep{GildepazApJS2007}, optical \citep{RC3} and NIR 
\citep{JarrettAJ2003}; dust emission has been observed at 12, 25, 60 
and 100~$\mu$m \citep{SandersAJ2003}, 170 and 200~$\mu$m 
\citep{PopescuA&A2003a}, 260, 360 and  580~$\mu$m \citep{DupacMNRAS2003}, 
450 and 850~$\mu$m \citep{AltonApJL1998} and 1.3~mm \citep{GuelinA&A1993}.
These fluxes are shown as datapoints in Fig.~\ref{allseds}.

\subsection{The SED of smooth disks}

The structural properties of NGC~891 have been determined in the
optical/NIR bands by \citet{XilourisSub1998} 
\citep[see also][]{XilourisA&A1998}. Surface brightness fits suggest
a thin stellar disk with $h_\mathrm{s}/z_\mathrm{s}\approx 10$, and a
radial scalelength $h_\mathrm{s}$ increasing from the K to the B-band
(a well known feature in less inclined object; see, e.g., 
\citealt{DeJongA&A1996b}). For simplicity, I neglect the gradient and
assume $h_\mathrm{s}=4$~kpc, a value appropriate for the NIR bands,
where the bulk of stellar emission is; $z_\mathrm{s}=0.4$~kpc is used 
for the vertical scalelength.

\begin{figure*}
\centering
\includegraphics[width=8.5cm]{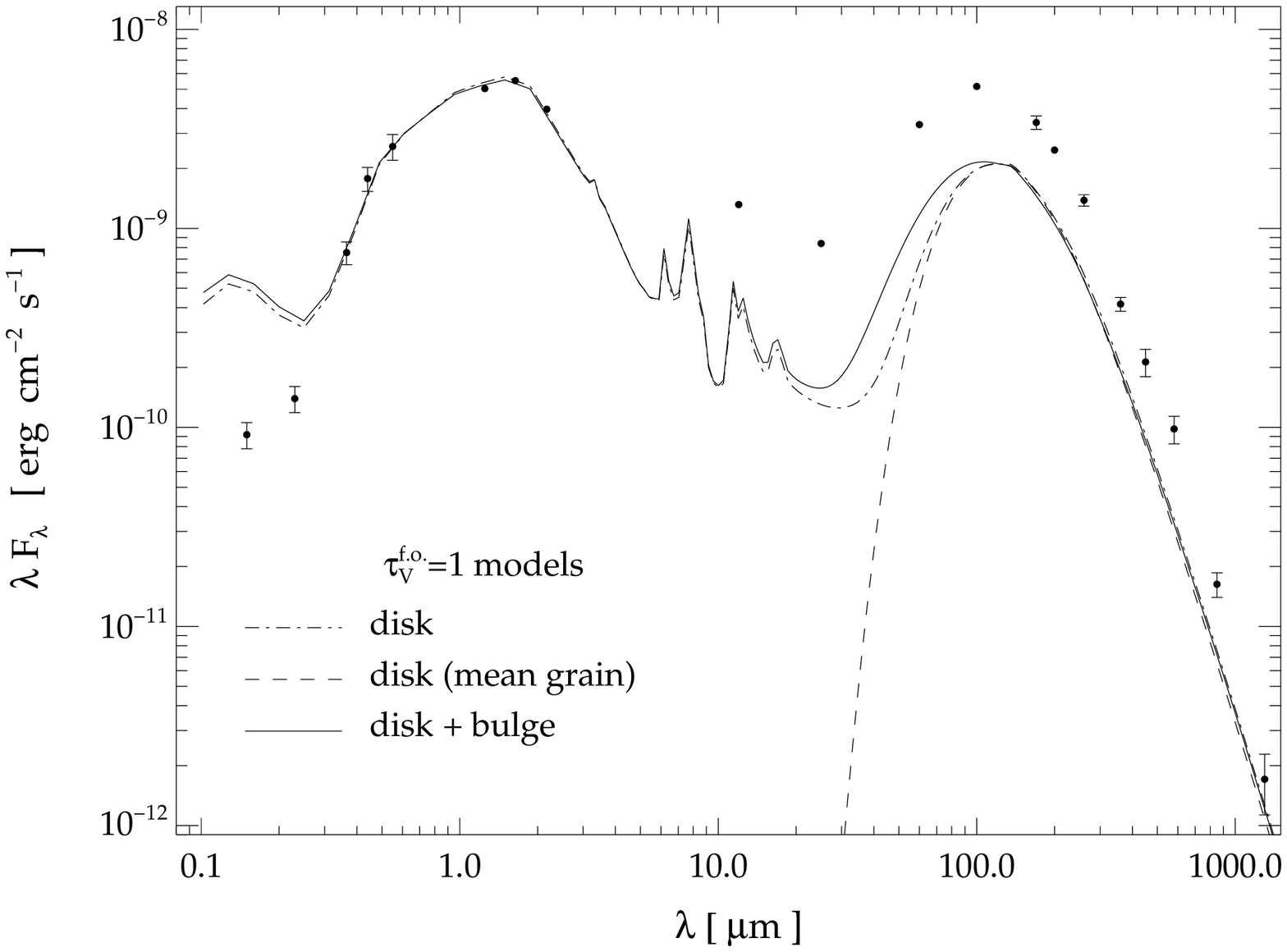}
\includegraphics[width=8.5cm]{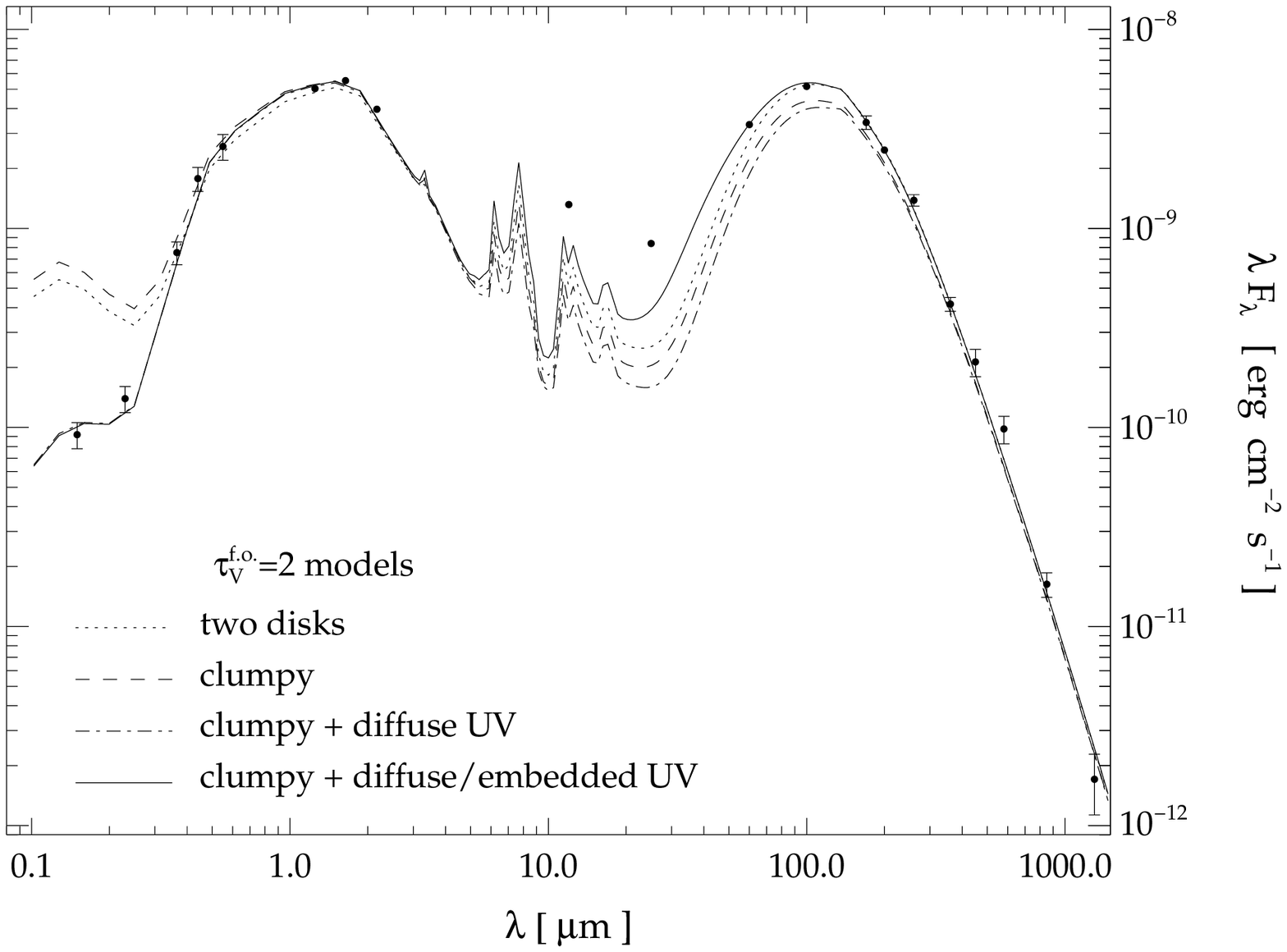}
\caption{SEDs for edge-on models. The left panel shows the models
for a $\tau^\mathrm{f. o.}_\mathrm{V}=1$ and $L_\mathrm{bol} = 
6\times 10^{10} L_{\sun}$. As TRADING computes separately dust extinction 
and emission, the total SEDs have been obtained by interpolating the two 
outputs over the common wavelength range and summing them together. 
Dust emission only is shown for the case of thermal equilibrium 
emission by a single grain with mean properties (dashed line), after
correcting for the energy absorbed by small grains.
The right panel shows the SEDs of $\tau^\mathrm{f. o.}_\mathrm{V}=2$
models. In all cases, the 
luminosity emitted by low mass stars is $6\times 10^{10} L_{\sun}$.
In both panels, data points represent the observed SED of NGC891 from 
literature (see text for references). UV, optical and NIR fluxes have been
corrected for foreground Galactic Extinction. When not shown,
error bars are smaller than symbol size.
}
\label{allseds}
\end{figure*}

The fits of \citet{XilourisSub1998} also provide a description of the
dust distribution. Since the radial scalelength of the dust disk almost 
doubles the NIR stellar scalelength, I have assumed $h_\mathrm{d}=2
h_\mathrm{s}$. The vertical scalelength is taken to be 
$z_\mathrm{d}=z_\mathrm{s}/2$, and $\tau^\mathrm{f. o.}_\mathrm{V}=$1
as suggested by the same fits. Similar properties are shown by dust
disks in other edge-on galaxies \citep{XilourisSub1998,BianchiA&A2007}.

The modelled SED are shown in Fig.~\ref{allseds} (left panel). The stellar
radiation is assumed to have a bolometric luminosity $L_\mathrm{bol} = 
6\times 10^{10} L_{\sun}$, so as to provide a match of the optical/NIR
observations.
The result for a single stellar disk is shown for a full calculation 
including stochastic heating (dot-dashed line) and for a thermal equilibrium
calculation using a single dust grain of mean absorption cross section
(dashed line). While the simpler calculation is not able to produce 
significant radiation at $\lambda<40\mu$m, the output at $\lambda\ga100\mu$m 
is consistent with that of the more complex case, clearly showing that most 
of radiation in the dust emission peak is due to thermal equilibrium heating.
It is necessary, however, to exclude from the equilibrium calculations
the amount of energy (30\% for the case shown in Fig.~\ref{allseds})
which is absorbed by smaller grains \citep{BianchiA&A2000b}. In the 
following, all calculations will be done for the complete dust model
of Sect.~\ref{dust}.

An important contribution to ISRF and to dust heating may come from a
galactic bulge. Following the analysis of \citet{XilourisSub1998},
I have included a Sersic $n=4$ bulge with $b/a=0.6$ and 
$R_\mathrm{e}=1$~kpc, with a luminosity $0.3 L_\mathrm{bol}$.
The SED for this model is shown in Fig.~\ref{allseds} (solid line in
left panel).
The central concentration of the bulge results in an increased temperature
for the larger grains: while in a single stellar disk model the larger
carbonaceous grains attain a maximum temperature of about 28K in the
center (25 K for silicates), in the model including a bulge the central
regions have graphite grains at 65 K (and silicates at 53 K). This
is evident in the shift of peak of thermal emission towards shorter 
wavelengths. Changes are instead negligible for the stellar SED.
The bulge/disk configuration will be used in the rest of the paper.

The $\tau^\mathrm{f. o.}_\mathrm{V}=1$ disk shown so far grossly
underestimate (by almost a factor two) the FIR output, as it is able
to absorb (and reemit) only about 18\% of the total energy. 
The discrepancy is the well known energy balance problem: a moderate 
optically thick dust disk as suggested by optical observations is not 
able to absorb enough energy to match the observed infrared output; 
absorption from a more opaque dust component is needed 
\citep{BianchiA&A2000b,PopescuA&A2000,MisiriotisA&A2001}. 
A $\tau^\mathrm{f. o.}_\mathrm{V}=2$ disk with $L_\mathrm{bol} 
\approx 7.5\times10^{10}$ L$_{\sun}$ (not shown) can reproduce 
better the dust emission, but is still 20\% less bright than
observations around the peak.
However, such high optical depths are not not compatible with 
observations \citep{XilourisSub1998,BianchiA&A2007}.

The SEDs also fail in reproducing the MIR radiation and show an excess 
of UV radiation. Although the second problem can be due to an overestimate 
of the UV component in the intrinsic SED, it is more likely the result of 
an underestimate of opacity: smooth distributions for dust and stars
cannot easily reproduce the localised extinction of young objects close 
to their dense, more opaque, birth environment. 

\subsection{The SED of clumpy disks}
\label{clumping}

Since the $\tau^\mathrm{f. o.}_\mathrm{V}=1$ disk of \citet{XilourisSub1998} 
fails in reproducing the FIR SED, \citet{PopescuA&A2000} have introduced
a second dust component, of mass similar to that of the first disk, but with 
a radial distribution equal to the stellar. This second, smooth, disk 
is devised to mimic the extra extinction in molecular clouds, while the first 
could be thought as associated to the atomic gas. Indeed, the Interstellar 
Medium (ISM) in NGC891 is made by a broad \ion{H}{i} distribution and a 
more centrally peaked 
molecular component \citep[see, e.g., ][]{AltonSub1999}. The distribution
of molecular gas is generally found to be close to the stellar 
\citep{ReganApJ2001}.  The molecular and atomic gas in NGC891 have similar 
masses \citep[in total $8 \times 10^9$~M$_{\sun}$, 70\%  of which residing 
in a thin disk;][]{SofuePASJ1993,OosterlooAJ2007}, in agreement also with 
the mean properties of Sb galaxies \citep{YoungARA&A1991}.  

In this paper, the extra mass component required to solve the energy
balance problem is added in the form of a distribution of clouds.
The clouds are distributed in an exponential 
disk of radial scalelength $h_\mathrm{s}$. The vertical scalelength is the 
same as for diffuse dust. The clumpy component has the same mass as the
diffuse dust ($f_\mathrm{c}=0.5$). The total dust mass in the grid is
$M_\mathrm{dust}= 10^{8} M_{\sun}$. For the assumed $G/D$, the dust mass
is roughly consistent with the gas mass derived from observations.

In the model, clouds are spherical, homogeneous and their opacity is 
computed from the gas surface density observed in GMC. Asssuming
$\Sigma = 100 $ M$_{\sun}$ pc$^{-2}$, 
a typical value for Local Group galaxies 
\citep{BlitzProc2007,RosolowskyApJ2003}, 
the optical depth along the diameter is $\tau_\mathrm{V}^\mathrm{diam} 
\approx 9.2$, regardless of the cloud dimensions (as follows from 
Eq.~\ref{kcloud}, using the dust properties of Sect.~\ref{dust}). 
According to \citet{RosolowskyPASP2005}, the largest clouds have a 
hydrogen mass $M_\mathrm{sup}$ = 10$^{6.5}$ M$_{\sun}$, which correspond 
to a radius $R\approx 100$pc. GMCs in the Milky Way and a few other Local Group 
galaxies show a power law distribution, described by Eq.~\ref{mass_spectrum} 
with $\gamma=1.7$ (with the notable exception of M33; \citealt{BlitzProc2007}). 
A value of $\gamma<2$ ensures that most of the molecular mass resides
in high mass clouds. 

Memory requirements may become heavy for the models of a clumpy disk,
when a reasonably large resolution is required for a large number of 
clouds\footnote{ 
Executing the dust emission part of TRADING on the clumpy model 
presented in this section requires about 1.6Gb of RAM on a 
64-bit CPU to store a grid made of 3.2$\times10^6$ cells (90\% of 
which are {\em leaf cells}).  The memory load is mostly due to 
the pointer-based structure of the binary tree. As an increase
in resolution results both in an increase of memory requirements 
and calculation time, TRADING efficiency will certainly 
benefit from parallelization with a distributed memory scheme.}.
In particular, a lower mass limit $M_\mathrm{inf}$ implies
a larger number of lower mass clouds. Several tests were carried out
with different values of $M_\mathrm{inf}$, spanning up to
two orders of magnitude in mass. The results were found
to be almost independent of the choice on $M_\mathrm{inf}$,
because of the constant optical depth and of the luminosity of 
embedded radiation, which was chosen to scale with the mass of the cloud. 
Thus, a single cloud mass is assumed ($M_\mathrm{inf}=M_\mathrm{sup}$).
Because of this choice, it is possible to force resolution down to 
sublevel $l=4$ inside clouds
(while the adopted $E$ value ensures subdivision only down to $l=3$).
In total, there are about 1650 clumps\footnote{
Simulations with the clumpy dust grid shown here take about 65 hours 
on a 2.2 GHz AMD Athlon 64 X2 Dual Core Processor 4200+; of these, 58 
hours are needed by the dust emission calculation. Maps of 512x512
pixels, covering all the dust distribution, are produced for three 
inclinations; $N_\mathrm{phot}= 4\times 10^7$ is sufficient for an 
adequate S/N of the maps. Self-absorption is included and iteration 
is stopped if the SED converge to within 5\%. 
For all the disk models presented here, only two iterations were necessary, 
with the SED actually converging to within a few {\em tenths} of percent.
Self-absorption attenuates dust emission in the SED for 
$\lambda\la10\mu$m, especially for the models seen edge-on.  However, 
it is found to have little influence on the global energy balance.}.

The case of a clumpy disk heated by diffuse radiation is shown
in the right panel of Fig.~\ref{allseds} (dashed line). The intrinsic
radiation emitted by the disk and the bulge is that for the Sb template
of Fig.~\ref{sb_sed}, and the bolometric luminosity
$L_\mathrm{bol}=7.3 \times 10^{10} L_{\sun}$. When compared to the model
in which all dust is in diffuse components (dotted line; analogous to 
the two dust disk model of \citealt{PopescuA&A2000}), the clumpy
model shows the usual reduction in the attenuation of starlight
\citep[see, e.g. ][]{WittApJ1996,VarosiPrep1999}. This
is reflected by a reduced dust emission in the infrared, for
$\lambda \la 200 \mu$m. At longer wavelengths, instead, the difference
between a clumpy and a smooth distribution is not appreciable.
Both diffuse dust and dust in clouds externally heated by the ISRF 
\citep[{\em passive} or {\em quiescent} clumps;][]{PopescuA&A2000} contribute 
to the FIR/submm emission. For the cloud optical depth considered here,
dust in {\em quiescent clumps} is not substantially colder than the
diffuse dust, and the spectral shape of the two components is similar.
The FIR/submm emission is thus proportional to the total dust mass in
a galaxy. 
The result holds even if $\Sigma$ (and thus the cloud optical depth)
is doubled: the diffuse NIR ISRF is still able to penetrate a cloud,
and the SED at longer wavelengths shows no appreciable change.
Similar conclusions are drawn by \citet{MisseltApJ2001}, though they
show a more pronounced dependence of the thermal peak on clumping.
The difference is probably due to the different heating geometry, with
a clumpy shell heated by an internal spherical distribution of stars.

Until now, all the stellar components have shared the same spectrum.
In the next model, the low-mass spectrum of Fig.~\ref{sb_sed} is
used for the disk and the bulge. These components have a total
luminosity $6 \times 10^{10} L_{\sun}$, with 70\% of the radiation emitted
by the disk. This is the same luminosity the old component has
in the clumpy model shown before, which is close to the observed
stellar SED in the optical and NIR.

Two UV emitting components are then included, sharing the high-mass
spectrum of Fig.~\ref{sb_sed}. The first component is a diffuse disk,
with the same radial and vertical scalelengths as the clump distribution.
After being extinguished by the adopted dust geometry, the spectrum
becomes flat, at least for the edge-on case. If such component, with
luminosity $0.8 \times 10^{10} L_{\sun}$, is added to the old stars,
the stellar SED is well matched also at UV wavelengths 
(Fig.~\ref{allseds}, right panel, dot-dashed line). The dust 
emission SED is close to that of the clumpy model with 
no separate emission from high-mass stars.

A second UV component is represented by point sources emitting radiation 
from the center of each cloud. This component of embedded sources is
heavily extinguished; stellar radiation at small wavelengths is almost 
completely absorbed and the output spectrum peaks at $1-2 \mu$m. Because of 
the high extinction, a considerable luminosity can be assigned to this 
component without
modifying the stellar SED of the simulation. The luminosity is instead
constrained by dust emission: the peak of thermal emission can be 
reproduced with $1.0 \times 10^{10} L_{\sun}$.  The final SED is shown 
with a solid line in Fig.~\ref{allseds} 
(right panel). There is a good accordance with observations along 
all the spectrum, with the exception of a lower MIR flux in the
simulation. I will comment on this later when discussing  the
dependence of the model on the chosen resolution. No significant
changes in the SED can be obtained if a fraction of the cloud is 
taken to be only {\em passive} (e.g.\ without sources inside);
if the UV source is positioned randomly in the cloud volume; if 
the UV luminosity emitted in a cloud is shared by a few sources.
If instead the radiation is emitted uniformly inside the cloud,
the SED peaks at colder wavelengths (an effect similar to a lack
of resolution; see later).
If $\Sigma$ is doubled, each cloud becomes smaller and more opaque:
the SED does not change significantly, with deviations within
5\% at the peak of the SED and 10\% at about 20$\mu$m.

In total, the model has $L_\mathrm{bol}=7.8 \times 10^{10} L_{\sun}$;
The UV-dominated spectrum of high-mass stars emits about 23\% of 
$L_\mathrm{bol}$ (sligthly more than in the Sb template), and 55\% of this 
radiation is embedded inside
clumps. By using standard calibrations \citep{KennicuttARA&A1998},
the intrinsic, dust-free, luminosity at 0.15 $\mu$m can be converted to 
a SFR$\approx 3 M_{\sun}$ yr$^{-1}$, a value similar to the UV component
used in the model of \citet{PopescuA&A2000}.
Of all the bolometric radiation, 34\% is absorbed globally; 20\% is
absorbed at $\lambda \ga 0.35\mu$m. Despite radiation from high-mass
stars suffers a larger internal extinction, still radiation from 
low-mass, older, stars dominate the dust heating. This is instead in
contrast with \citeauthor{PopescuA&A2000}, where UV radiation dominates
the energy balance, contributing to 70\% of the FIR radiation.
The reason for this discrepancy is unknown. Since the derived SFRs
are similar, it should be due to model differences in the optical/NIR.
In the current model,
extinction of optical radiation could have been overestimated
at around $\lambda = 0.5\mu$m, since the stellar scalelength
is taken to be about 30\% smaller than what deduced from 
B and V-band observations \citep{XilourisSub1998}; this makes
the stellar disk more concentrated with respect to the diffuse 
dust. However, most of observed radiation comes at longer wavelengths, 
where the adopted value for $h_\mathrm{s}$ is closer to the fits. 
Furthermore, the second dust disk of \citet{PopescuA&A2000} is
likely to absorb more radiation than the clumpy component in this
work.

The current model also differs in the fraction of UV radiation which
is embedded in clouds. In \citet{PopescuA&A2000}, dust emission caused 
by absorption of UV photons from localised sources is modelled using
a SED template based on observations of a Galactic \ion{H}{ii} region. This
template consists in a modified blackbody with emissivity 
$\propto \lambda^{-2}$ and $T\approx 35$~K; it does not include, 
in the words of the authors, {\em potentially cold emission components
that might be expected from "parent" molecular clouds in juxtaposition 
to their "offspring" \ion{H}{ii} regions}. From the point of view of the
current work, they supply a {\em potentially cold emission}
component by adding the second dust disk. 
However, in \citet{PopescuA&A2000} the calculation for localised emission
is not done self-consistently with that of the second, diffuse, disk.
This might have altered the balance of the UV components, and be the
cause of the discrepancy with the results shown here.

\citet{DasyraA&A2005} use the \citet{XilourisSub1998} disk and increase 
the dust emissivity, without adding a second disk. As a result, the dust 
temperature reduces and the thermal emission SED shifts toward longer 
wavelengths, thus matching the FIR/submm observations. The missing flux at 
lower wavelengths is provided by absorption of UV starlight with luminosity 
(SFR) larger than in \citet{PopescuA&A2000}. Indeed, observations and 
models suggest that the dust emissivity could be larger, at least in 
denser environment \citep[see, e.g., the discussion in ][]{AltonA&A2004}. 
However, the larger dust emissivity they derive may be simply due to the 
lack of cold dust emission in the \ion{H}{ii} region template 
they use \citep[which is the same as in][]{PopescuA&A2000}. 

\begin{figure}
\resizebox{\hsize}{!}{ \includegraphics{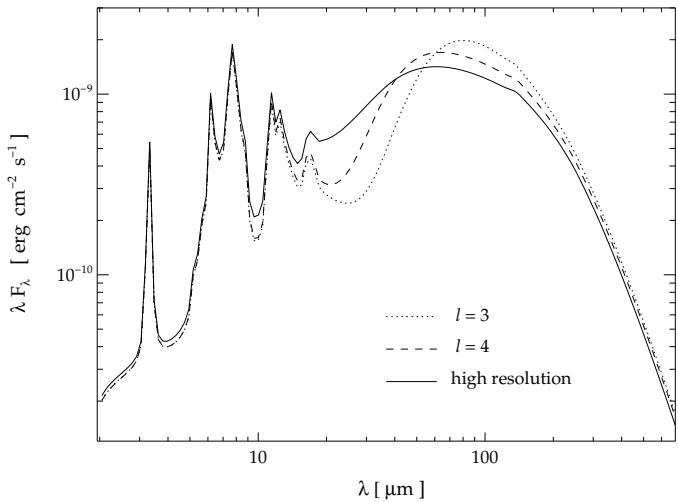} }
\caption{
Grid resolution convergence check for the emission from clouds in the
clumpy model. Only dust emission is shown (see text for details).
}
\label{sed_res}
\end{figure}

For the clumpy grid shown here and with the current version of TRADING, 
it is not possible to check for the convergence of the results by simply
increasing the resolution within clouds. However, a test can be made by
comparing: a simulation in which only the embedded UV radiation is
emitted from the center of the clouds; a simulation of a single cloud 
with the high resolution grid used for the benchmark cases 
(Sect.~\ref{bench}), scaled to the total number of clouds. This
is shown if Fig.~\ref{sed_res} (dust emission only). If the cloud 
resolution is limited to sublevel $l=3$ (dotted line), the peak emission 
is at larger wavelengths than in the high resolution calculation (solid line).
The situation improves by forcing resolution to sublevel $l=4$ (dashed line; 
the value used in this paper),
since larger thermal equilibrium temperatures can be sampled by
a higher spatial resolution closer to the embedded source. Because 
of the high extinction in the UV, the central source does not heat 
significantly the external parts of the cloud. In full simulation 
these are predominantly heated by the diffuse ISRF, for which
a coarser cloud and smooth medium resolution is sufficient. Thus,
the differences around the peak emission are smaller than in 
Fig.~\ref{sed_res}. This was confirmed by tests on models with
less memory requirements (smaller number of clouds).
For the continuum emission at shorter wavelengths,
however, differences are higher and could explain the low
flux predicted at $24\mu$m in the SEDs of Fig.~\ref{allseds}.

Instead, the PAH features show a negligible change with resolution, 
since they depend only linearly, for the temperature 
range considered here, on the intensity of the ISRF (see, e.g.,
Fig.~\ref{isrf} and \citealt{DraineApJ2007b}). The observed 
flux at $12\mu$m in Fig.~\ref{allseds} could be matched by increasing 
the amount of embedded radiation of about a factor two; however, this 
would also shift the peak of the simulated SED at shorter wavelengths.
Also, the flux cannot be increased by increasing the amount of 
mass in clouds (and thus their number), since the mass is
constrained by the FIR/submm observations.
Thus, the low predicted flux in the PAH features may point
towards different ISRF intensities than what can be obtained 
here with a simple cloud description.

The assumption of spherical, homogeneous, clouds is clearly
a poor description for the complex structure of the molecular 
medium, which appears to be fractal \citep{BlitzProc1999}.
Unfortunately, the current, non-parallel, version of TRADING 
and the limitations in the available computer memory do not 
allow a detailed study of the effects of such structure on
the models output. Limited tests have been carried out by
assuming that only a fraction of the cells within the spherical 
volume of each cloud is occupied by dust associated with 
molecular gas, while the remaining cells have the same properties 
of the underlying smooth disk. I have run two cases with filling 
factors $ff=0.5$ and $0.2$. In the first case, clouds maintain
a structural integrity, being made by many interconnected cells,
while in the second the structure is looser \citep{MisseltApJ2001}.
The density of the dust associated with molecules increases by
a factor $1/ff$ with respect to the standard models, and this
already may pose resolution problems, expecially for the $ff=0.2$
case if the central source happen to be within a higher density 
cell. As expected, more UV radiation is able to escape clouds in 
these cases (though the $ff=0.5$ case is still consistent with 
observations while the $ff=0.2$ is only higher by about 30\%).
As a consequence of that (and probably of the limited resolution)
the flux between 20 and 80 $\mu$m is lower than in the standard
model, by 20\% and 40\% at 30$\mu$m for the $ff=0.5$ and $0.2$ 
cases, respectively. However, there are no significant differences 
in the optical SED (because emission comes prevalently from the 
diffuse medium), in the PAHs range and in the thermal peak at
larger wavelengths.

\subsection{Attenuation {\em vs} inclination and images}

\begin{figure*}
\centering
\includegraphics[width=16cm]{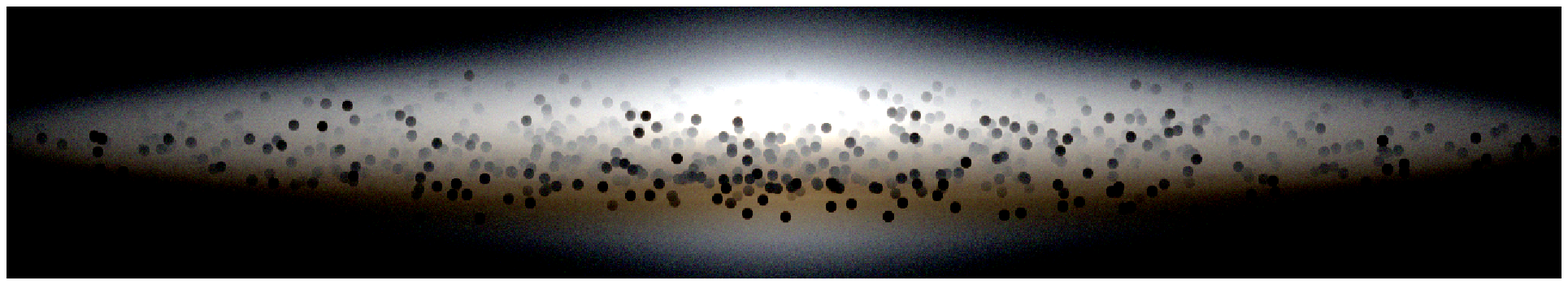}
\includegraphics[width=16cm]{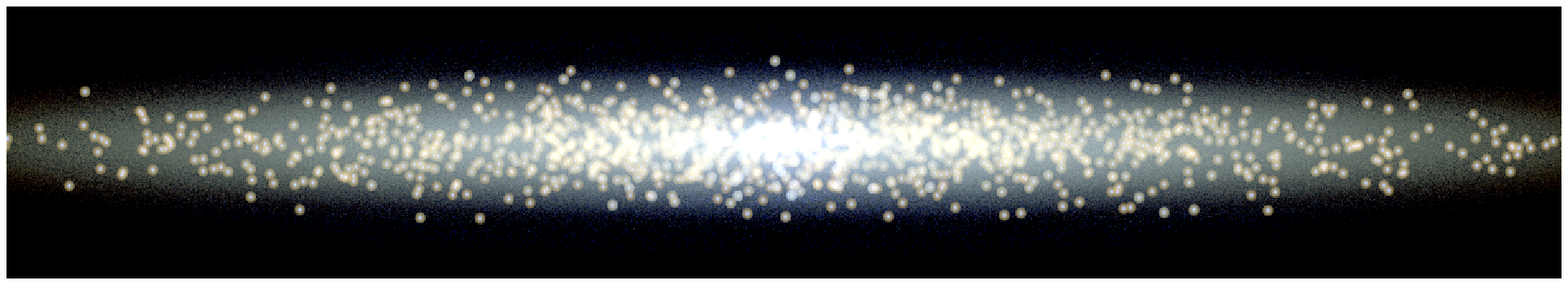}
\caption{RGB true color images of the clumpy model, seen from $i=86^\circ$.
The upper panel shows an image in the optical (Red = K-band, Green= V-band,
Blue= B-band). The lower panel shows an image in the FIR/submm, around the 
peak of dust emission (Red = 850 $\mu$m, Green= 250 $\mu$m, Blue= 70 $\mu$m).
Each panel has an extent of 32 kpc x 3.5 kpc.
}
\label{truec}
\end{figure*}

The global SED shown before have been obtained by integrating over
simulated images for the edge-on case. However, images can be produced
at any inclination. In Fig.~\ref{truec}, true color images are shown
for the clumpy model with embedded sources, seen at $i=86^\circ$. 
The upper panel shows an image in the optical. Reddening is evident
in the extinction lane and is mostly due to the diffuse disk, while
clumps, due to their high optical depth, simply extinguish all radiation 
in their background, regardless of the wavelength. The lower panel
shows an image in the FIR/submm around the peak of dust emission.
Dust in clumps is hotter (bluer) in their centers, due to the embedded
sources, and colder (redder) close to their surface, where the heating
is mostly due to the diffuse ISRF. Emission from diffuse dust shows a 
gradient, which is strong only in the center of the disk, where the
dust temperature is larger (mostly as an effect of the stronger
ISRF in the bulge region).

\begin{figure}
\resizebox{\hsize}{!}{ \includegraphics{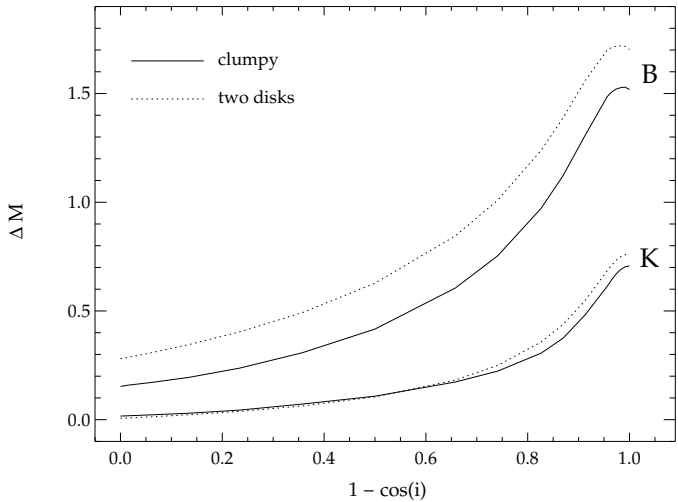} }
\caption{Attenuation as a function of inclination. $\Delta M$ is the 
difference, in magnitude, between the attenuated flux and the flux that 
would be emitted in the absence of dust. }
\label{incli}
\end{figure}

The two dust disks model of \citeauthor{PopescuA&A2000} 
(\citeyear{PopescuA&A2000}; see also \citealt{TuffsA&A2004}) has
been successful in reproducing the dependence of attenuation
on inclination, as obtained from the analysis of the luminosity
function of disk galaxies with different axis ratio in the 
Millennium Galaxy Catalogue \citep{DriverMNRAS2007}. In 
Fig.~\ref{incli}, I show the difference (in magnitude) between
the total magnitude and the magnitude in the no-dust case (the 
latter being independent of $i$ for the isotropic emission assumed 
here). In the B-band, attenuation for the clumpy model is reduced,
as expected, with respect to the two disks model\footnote{
I note here that the two disks model shown in this section and
in Fig.~\ref{allseds}, though analogous in concept, {\em is not the 
same} as the model of \citet{PopescuA&A2000}. In this paper, it is 
used only to show the changes when passing from a diffuse distribution
to a clumpy distribution for the dust component associated to the 
molecular gas. The main difference is in the radial scalelength adopted 
for this component. In \citet{PopescuA&A2000} the scalelength is larger, 
resulting, for a similar dust mass, in a disk with a face-on optical 
depth roughly half of that for the molecular component in this work.
}.
However, the dependence 
on $i$ is rather similar. Thus, the clumpy model could in principle 
be consistent with observations, though a more detailed study
is needed, with the separate analysis of the attenuation of the
stellar bulge and disk. The reduction in attenuation (at least 
in the B-band) does not seem large enough to substantially alter 
the conclusions drawn by \citet{DriverApJL2008} on the energy 
balance of the Cosmic SED.

\begin{figure}
\resizebox{\hsize}{!}{ \includegraphics{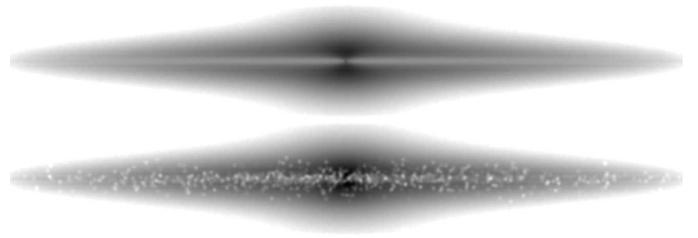} }
\caption{K-band images for the two dust disks (top) and 
clumpy (bottom) models, seen edge-on. The extent of the
images along the major axis is 32 kpc.}
\label{images}
\end{figure}

In the K-band, the attenuation is rather similar both in amount
and dependence on $i$. Indeed, when seen edge-on, both models show
a dust lane of similar depth (Fig.~\ref{images}). In the two dust
disks model, most of extinction is due to the more centrally 
concentrated second dust disk, which has a higher opacity 
($\tau^\mathrm{f. o.}_\mathrm{V}\approx 4$). In the clumpy model, it 
is due to clouds,
which still have a non negligible optical depth ($\tau\approx 1$)
and, acting more like foreground dust, are more efficient in 
extinguishing radiation. Using the procedure of \citet{BianchiA&A2007},
the clumpy image could be fitted with a smooth disk with  
$\tau^\mathrm{f. o.}_\mathrm{K}\approx 0.4$ (the fitted radial
scalelength is intermediate between that of clumps and of the diffuse disk,
while the vertical scalelength is retrieved correctly). This is at odds 
with observations, as NGC~891 and other edge-ons do not show a pronounced 
extinction lane and have $\tau^\mathrm{f. o.}_\mathrm{K}\la 0.1$
\citep{XilourisSub1998,DasyraA&A2005,BianchiA&A2007}. A fit to a
simulated image in the V-band, instead, retrieves the parameter of the 
diffuse disk: at optical wavelengths this structure is more evident than 
in the NIR and coherent while the clumps behaves as {\em noisy} spots that
increase the fit residuals (analogous to the inhomogeneities in real images).
In optical wavelengths, thus, it is possible to {\em hide} dust in clumps,
though the underestimation depends on the model assumptions
\citep{BianchiSub1999,MisiriotisA&A2002}.

\begin{figure}
\resizebox{\hsize}{!}{ \includegraphics{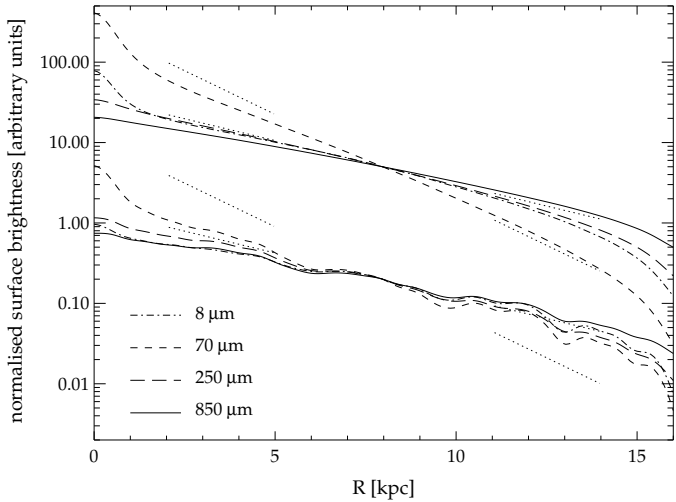} }
\caption{Major axis surface brightness profiles for the two
disks (top) and clumpy (bottom) models, seen edge-on. The
profiles have been convolved with a gaussian PSF (FWHM=16'');
smoothed by folding the left and right part of the image
(for the clumpy model); arbitrarily normalised at 8kpc.
As an aid to the eye, the dotted segments show the slopes
of exponential fall-offs with scalelengths $h_\mathrm{s}
=4$kpc and $h_\mathrm{s}/2$.
}
\label{radial}
\end{figure}

In Fig.~\ref{radial}, I show the major axis surface brightness
profiles of dust emission for the two disks and the clumpy 
models seen edge-on. Images have been smoothed with a PSF of 
FWHM=16'', a resolution similar to what can be achieved, currently or in
the next years,  at 70$\mu$m using MIPS aboard the satellite 
Spitzer \citep{RiekeApJS2004}, at 250$\mu$m using SPIRE
aboard the satellite Herschel \citep{GriffinAdSpR2007} and at 
850$\mu$m using the submm cameras SCUBA2 \citep{HollandProc2006} 
and LABOCA \citep{KreysaProc2003}. To these, I have added a
profile at about 8$\mu$m, centered on a large PAH feature.
For a diffuse dust disk co-spatial with the heating sources,
as is the case in the two dust disk model for the second dust 
disk and the stellar disk, the emission gradient becomes less steep
as the wavelength increases. In the submm, the Rayleigh-Jeans spectrum 
does not depend strongly on the dust temperature, and the gradient
approaches the intrinsic slope of the dust distribution (close to the 
radial scalelength $h_\mathrm{s}$\footnote{The radial profile at 
850$\mu$m is steeper than the exponential of scalelength $h_\mathrm{s}$, 
both because of the contribution of the diffuse disk, and because the 
projected profile of an exponential disk seen edge-on is 
$\propto r K_1(r/h)$, with $K_1(x)$ the modified Bessel function of the 
second kind and first order and $r$ the distance from the center 
\citep{KylafisApJ1987}.  For an infinite disk, the profile tends 
to the exponential. In the case shown here, the second dust disk, 
the distribution of clumps, and the stellar disk are all 
truncated at 4 $h_\mathrm{s}$. This causes the profile steepening 
seen at 16~kpc. The truncation has little effect on the results.}). 
Instead, at wavelengths shorter than the peak, the spectrum depends
more on the strength of the ISRF, as shown by the 70$\mu$m profile
(the further steepening in the center being due to the bulge).
This is not the case for the PAH emission, however, 
because of the linear dependence of the spectrum on the ISRF intensity.

When the dust mass of the second disk is distributed in clumps (retaining
the same radial scalelength), the differences between the profiles at
all wavelengths are reduced. This is caused mainly by the contribution
of internal clump emission at $\lambda < 200\mu$m, which is independent 
of the clump position. Also, the $8\mu$m profile becomes closer to the 
submm, explaining the observed correlation between PAHs and cold dust
\citep[see, e.g.][]{HaasA&A2002}. Though reduced, the bulge contribution 
to the 70$\mu$m emission is still detectable (see also Fig.~\ref{truec}).

\begin{figure}
\resizebox{\hsize}{!}{ \includegraphics{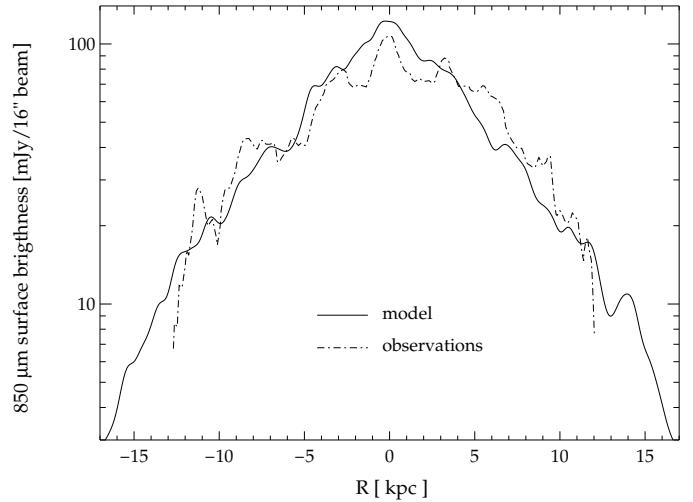} }
\caption{Major axis surface brightness profile at 850 $\mu$m
for the clumpy model, compared to the SCUBA observations of 
\citet{AltonSub1999}. The model has been convolved to the
SCUBA beam (FWHM=16'') and the profile smoothed as in 
\citet{AltonSub1999}.
}
\label{scuba}
\end{figure}

Finally, I show in Fig.~\ref{scuba} a comparison between the
clumpy model and the SCUBA major axis profile as presented
in \citet{AltonSub1999}. As for model of \citet{PopescuA&A2000}
(see their Fig.~6) there is a 
broad agreement with observations, both in flux level and 
gradient. Only, the clumpy model appears to be {\em smoother} 
than observations, possibly hinting to a more complex structure
in the cloud distribution than what adopted for this paper.

\section{Summary}
\label{summary}

I have presented in this paper the Monte Carlo radiative transfer 
code TRADING. It is designed to study continuum extinction and 
emission in a dusty medium. The main features of TRADING are:
\renewcommand\theenumi{\roman{enumi}}
\begin{enumerate}
\item a description of the dust distribution based on a binary-tree 
adaptive grid;

\item possibility of emitting stellar radiation from several
diffuse distributions and collections of point sources/clouds, 
each with an independent spectrum;

\item dust emission from a distribution of grains of different sizes
and materials, under thermal equilibrium or stochastic heating 
conditions;

\item self absorption through an iterative process.

\end{enumerate}

The code shows an excellent agreement with benchmark cases and
can be easily adapted to several environments of astrophysical 
interest. For the application shown in this paper, TRADING has 
been tailored to the case of a galactic disk. The disk includes:
a diffuse dust component; a clumpy component made of a distribution
of spherical clouds, with properties modelled on those of GMC in 
local galaxies. Building on the studies of \citet{XilourisSub1998}
and \citet{PopescuA&A2000}, simulations have been produced for
the well known case of the edge-on spiral NGC~891. The main results
can be summarised as follows.

\renewcommand\theenumi{\arabic{enumi}}
\begin{enumerate}
\item The global SED of NGC~891 from the UV to the mm can be reproduced
by a model including: a diffuse dust disk, more widespread than the
stellar and of moderate optical depth ($\tau^\mathrm{f. o.}_\mathrm{V}$=1);
a clumpy dust distribution, of mass similar to that of the diffuse disk; 
a population of old stars, smoothly distributed in an exponential disk 
and a bulge, emitting about 75\% of the bolometric radiation; a population
of young stars emitting the remaining fraction mainly in the UV,
roughly half of which embedded in clouds. Despite the increased attenuation
for UV radiation, the old stars are the main contributor to dust heating
(about 60\% of all radiation emitted by dust). 
This is in contrast with the result of \citet{PopescuA&A2000}, which predict
a dominant UV contribution to heating.

In the MIR, the current model
produces less radiation than the observations. This can be only in part
explained with a lack of spatial resolution for clouds. The difference
can be possibly due to a more complex structure of the ISM and ISRF than 
modelled here. This could also be the reason for the larger extinction in 
the K-band image and the smoother appearence of the 850$\mu$m than observed.

\item Dust in clouds is found to be significantly heated by both the 
embedded UV radiation and the NIR radiation from the diffuse stellar 
distribution penetrating the outer shells of a cloud. 
This second source of heating results in cold dust emission which
is not present in the template for localised emission used by 
\citet{PopescuA&A2000} and \citet{DasyraA&A2005}.
In \citet{PopescuA&A2000}, cold dust emission is provided by a 
second diffuse dust disk; however, contrary to this work, the radiative 
transfer calculation for the extra dust component in
\citet{PopescuA&A2000} is not done self-consistently with that for 
localised sources. This is the cause of the larger fraction of embedded
UV radiation in the current work. Also, the difference between a
clumpy and smooth disk results in a smaller attenuation of diffuse
radiation at optical wavelengths in the current model.
In \citet{DasyraA&A2005}, the lack of cold dust emission associated
with localised emission appears to be the main reason of their claim 
for a larger dust emissivity in the FIR.
Other models which compute the contribution of clouds independently 
from that of the diffuse ISM, such as GRASIL \citep{SilvaApJprep1998},
might need to include external heating of clouds by a diffuse ISRF.
Because of the diffuse heating, cold emission in the submm comes 
from all the dust components.

\item In a clumpy disk, and for the resolution of current and
future instruments, color gradients are rather similar along all
the dust emission spectrum. They follow the gradients of stars
and of the main dust component. A steepening may be observed
in the Wien side of the spectrum for thermal equilibrium heated 
grains (20$\mu$m $\la \lambda \la$ 100$\mu$m), mostly because of
the presence of a bulge.

\end{enumerate}

The conclusions drawn on the energy budget for the single object 
NGC~891 might not be of general application for other galaxies,
though NGC~891 is considered to be a prototypical spiral.
Unfortunately, not many galaxies have detailed observations of
the FIR/submm beyond the thermal peak. However, the advent of the 
SPIRE instrument aboard the Herschel satellite \citep{GriffinAdSpR2007},
in conjunction with the new ground-based submm facilities, will soon 
permit a more detailed study of this spectral range, necessary 
for constraining the bulk of dust mass. 

\begin{acknowledgements}
I am grateful to 
Maarten Baes, 
Bruce Draine, 
Edvige Corbelli, 
Karl Gordon, 
Cristina Popescu,
Malcolm Walmsley
and
Emmanuel Xilouris
for helpful suggestions and discussions. 
I thank an anonymous referee for constructive comments
that improved the contents and presentation of this paper.
This research has made use of the NASA/IPAC Extragalactic Database 
(NED) which is operated by the Jet Propulsion Laboratory, California 
Institute of Technology, under contract with the National Aeronautics 
and Space Administration. This research has made use of NASA's 
Astrophysics Data System.

This paper is dedicated to the memory of Angelos Misiriotis,
a fond radiative transfer friend who passed away too prematurely.
\end{acknowledgements}

\bibliographystyle{aa}
\bibliography{/home/tinia/sbianchi/tex/DUST}

\begin{thebibliography}{86}
\expandafter\ifx\csname natexlab\endcsname\relax\def\natexlab#1{#1}\fi

\bibitem[{{Alton} {et~al.}(1998){Alton}, {Bianchi}, {Rand}, {Xilouris},
  {Davies}, \& {Trewhella}}]{AltonApJL1998}
{Alton}, P.~B., {Bianchi}, S., {Rand}, R.~J., {et~al.} 1998, ApJ, 507, L125

\bibitem[{{Alton} {et~al.}(2000){Alton}, {Xilouris}, {Bianchi}, , {Davies}, \&
  {Kylafis}}]{AltonSub1999}
{Alton}, P.~B., {Xilouris}, E.~M., {Bianchi}, S., {et~al.} 2000, A\&A, 356, 795

\bibitem[{{Alton} {et~al.}(2004){Alton}, {Xilouris}, {Misiriotis}, {Dasyra}, \&
  {Dumke}}]{AltonA&A2004}
{Alton}, P.~B., {Xilouris}, E.~M., {Misiriotis}, A., {Dasyra}, K.~M., \&
  {Dumke}, M. 2004, A\&A, 425, 109

\bibitem[{{Baes} {et~al.}(2003){Baes}, {Davies}, {Dejonghe}, {Sabatini},
  {Roberts}, {Evans}, {Linder}, {Smith}, \& {de Blok}}]{BaesMNRAS2003}
{Baes}, M., {Davies}, J.~I., {Dejonghe}, H., {et~al.} 2003, MNRAS, 1081

\bibitem[{{Baes} \& {Dejonghe}(2001)}]{BaesMNRAS2001a}
{Baes}, M. \& {Dejonghe}, H. 2001, MNRAS, 326, 722

\bibitem[{{Baes} {et~al.}(2004){Baes}, {Dejonghe}, \& {Davies}}]{BaesProc2004}
{Baes}, M., {Dejonghe}, H., \& {Davies}, J.~I. 2004, in IAU Symposium, Vol.
  220, Dark Matter in Galaxies, ed. S.~{Ryder}, D.~{Pisano}, M.~{Walker}, \&
  K.~{Freeman}, 343

\bibitem[{{Baes} {et~al.}(2005){Baes}, {Stamatellos}, {Davies}, {Whitworth},
  {Sabatini}, {Roberts}, {Linder}, \& {Evans}}]{BaesNewA2005}
{Baes}, M., {Stamatellos}, D., {Davies}, J.~I., {et~al.} 2005, New Astronomy,
  10, 523

\bibitem[{{Bianchi}(2007)}]{BianchiA&A2007}
{Bianchi}, S. 2007, A\&A, 471, 765

\bibitem[{{Bianchi} {et~al.}(2000{\natexlab{a}}){Bianchi}, {Davies}, \&
  {Alton}}]{BianchiA&A2000b}
{Bianchi}, S., {Davies}, J.~I., \& {Alton}, P.~B. 2000{\natexlab{a}}, A\&A,
  359, 65

\bibitem[{{Bianchi} {et~al.}(2000{\natexlab{b}}){Bianchi}, {Ferrara}, {Davies},
  \& {Alton}}]{BianchiSub1999}
{Bianchi}, S., {Ferrara}, A., {Davies}, J.~I., \& {Alton}, P.~B.
  2000{\natexlab{b}}, MNRAS, 311, 601

\bibitem[{{Bianchi} {et~al.}(1996){Bianchi}, {Ferrara}, \&
  {Giovanardi}}]{BianchiApJ1996}
{Bianchi}, S., {Ferrara}, A., \& {Giovanardi}, C. 1996, ApJ, 465, 127

\bibitem[{{Bjorkman} \& {Wood}(2001)}]{BjorkmanApJ2001}
{Bjorkman}, J.~E. \& {Wood}, K. 2001, ApJ, 554, 615

\bibitem[{{Blitz} {et~al.}(2007){Blitz}, {Fukui}, {Kawamura}, {Leroy},
  {Mizuno}, \& {Rosolowsky}}]{BlitzProc2007}
{Blitz}, L., {Fukui}, Y., {Kawamura}, A., {et~al.} 2007, in Protostars and
  Planets V, ed. B.~{Reipurth}, D.~{Jewitt}, \& K.~{Keil}, 81--96

\bibitem[{{Blitz} \& {Williams}(1999)}]{BlitzProc1999}
{Blitz}, L. \& {Williams}, J.~P. 1999, in NATO ASIC Proc. 540: The Origin of
  Stars and Planetary Systems, 3

\bibitem[{{Bohren} \& {Huffman}(1983)}]{BohrenBook1983}
{Bohren}, C.~F. \& {Huffman}, D.~R. 1983, {Absorption and Scattering of Light
  by Small Particles} (New York: {Wiley})

\bibitem[{{Cashwell} \& {Everett}(1959)}]{CashwellBook1959}
{Cashwell}, E.~D. \& {Everett}, C.~J. 1959, {A Practical Manual on the Monte
  Carlo Method for Random Walk Problems} (New York: {Pergamos})

\bibitem[{{Dasyra} {et~al.}(2005){Dasyra}, {Xilouris}, {Misiriotis}, \&
  {Kylafis}}]{DasyraA&A2005}
{Dasyra}, K.~M., {Xilouris}, E.~M., {Misiriotis}, A., \& {Kylafis}, N.~D. 2005,
  A\&A, 437, 447

\bibitem[{{de Jong}(1996)}]{DeJongA&A1996b}
{de Jong}, R. 1996, A\&A, 313, 377

\bibitem[{{de Vaucouleurs}(1959)}]{DeVaucouleursBook1959}
{de Vaucouleurs}, G. 1959, Handbuck der Physik, ed. S.~{Flugge}, Vol.~53
  (Berlin: Springer), 275

\bibitem[{{de Vaucouleurs} {et~al.}(1991){de Vaucouleurs}, {de Vaucouleurs},
  {Corwin}, {Buta}, {Paturel}, \& {Fouque}}]{RC3}
{de Vaucouleurs}, G., {de Vaucouleurs}, A., {Corwin}, Herold~G., J., {et~al.}
  1991, Third Reference Catalogue of Bright Galaxies (Berlin: Cambridge
  University Press), {RC3}

\bibitem[{{Draine}(2003{\natexlab{a}})}]{DraineARA&A2003}
{Draine}, B.~T. 2003{\natexlab{a}}, ARA\&A, 41, 241

\bibitem[{{Draine}(2003{\natexlab{b}})}]{DraineApJ2003}
{Draine}, B.~T. 2003{\natexlab{b}}, ApJ, 598, 1017

\bibitem[{{Draine} \& {Li}(2001)}]{DraineApJ2001}
{Draine}, B.~T. \& {Li}, A. 2001, ApJ, 551, 807

\bibitem[{{Draine} \& {Li}(2007)}]{DraineApJ2007b}
{Draine}, B.~T. \& {Li}, A. 2007, ApJ, 657, 810

\bibitem[{{Driver} {et~al.}(2008){Driver}, {Popescu}, {Tuffs}, {Graham},
  {Liske}, \& {Baldry}}]{DriverApJL2008}
{Driver}, S.~P., {Popescu}, C.~C., {Tuffs}, R.~J., {et~al.} 2008, ApJL, 803

\bibitem[{{Driver} {et~al.}(2007){Driver}, {Popescu}, {Tuffs}, {Liske},
  {Graham}, {Allen}, \& {de Propris}}]{DriverMNRAS2007}
{Driver}, S.~P., {Popescu}, C.~C., {Tuffs}, R.~J., {et~al.} 2007, MNRAS, 379,
  1022

\bibitem[{{Dullemond} \& {Turolla}(2000)}]{DullemondA&A2000}
{Dullemond}, C.~P. \& {Turolla}, R. 2000, A\&A, 360, 1187

\bibitem[{{Dupac} {et~al.}(2003){Dupac}, {del Burgo}, {Bernard}, {Giard},
  {Lamarre}, {Laureijs}, {Pajot}, {Ristorcelli}, {Serra}, {Tauber}, \&
  {Torre}}]{DupacMNRAS2003}
{Dupac}, X., {del Burgo}, C., {Bernard}, J.~., {et~al.} 2003, MNRAS, 344, 105

\bibitem[{{Fioc} \& {Rocca-Volmerange}(1997)}]{FiocA&A1997}
{Fioc}, M. \& {Rocca-Volmerange}, B. 1997, A\&A, 326, 950

\bibitem[{{Frisken} \& {Perry}(2002)}]{FriskenGraphics2002}
{Frisken}, S.~F. \& {Perry}, R. 2002, Journal of Graphics Tools, 7, 1

\bibitem[{{Gil de Paz} {et~al.}(2007){Gil de Paz}, {Boissier}, {Madore},
  {Seibert}, {Joe}, {Boselli}, {Wyder}, {Thilker}, {Bianchi}, {Rey}, {Rich},
  {Barlow}, {Conrow}, {Forster}, {Friedman}, {Martin}, {Morrissey}, {Neff},
  {Schiminovich}, {Small}, {Donas}, {Heckman}, {Lee}, {Milliard}, {Szalay}, \&
  {Yi}}]{GildepazApJS2007}
{Gil de Paz}, A., {Boissier}, S., {Madore}, B.~F., {et~al.} 2007, ApJS, 173,
  185

\bibitem[{{Gordon} {et~al.}(2001){Gordon}, {Misselt}, {Witt}, \&
  {Clayton}}]{GordonApJ2001}
{Gordon}, K.~D., {Misselt}, K.~A., {Witt}, A.~N., \& {Clayton}, G.~C. 2001,
  ApJ, 551, 269

\bibitem[{{Griffin} {et~al.}(2007){Griffin}, {Abergel}, {Ade}, {Andr{\'e}},
  {Baluteau}, {Bock}, {Franceschini}, {Gear}, {Glenn}, {Griffin}, {King},
  {Lellouch}, {Madden}, {Naylor}, {Oliver}, {Olofsson}, {Page},
  {Perez-Fournon}, {Rowan-Robinson}, {Saraceno}, {Sawyer}, {Swinyard},
  {Vigroux}, {Wright}, \& {the SPIRE Consortium}}]{GriffinAdSpR2007}
{Griffin}, M., {Abergel}, A., {Ade}, P., {et~al.} 2007, Advances in Space
  Research, 40, 612

\bibitem[{{Gu\'elin} {et~al.}(1993){Gu\'elin}, {Zylka}, {Mezger}, {Haslam},
  {Kreysa}, {Lemke}, \& {Sievers}}]{GuelinA&A1993}
{Gu\'elin}, M., {Zylka}, R., {Mezger}, P.~G., {et~al.} 1993, A\&A, 279, L37

\bibitem[{{Guhathakurta} \& {Draine}(1989)}]{GuhathakurtaApJ1989}
{Guhathakurta}, P. \& {Draine}, B.~T. 1989, ApJ, 345, 230

\bibitem[{{Haas} {et~al.}(2002){Haas}, {Klaas}, \& {Bianchi}}]{HaasA&A2002}
{Haas}, M., {Klaas}, U., \& {Bianchi}, S. 2002, A\&A, 385, L23

\bibitem[{{Henyey} \& {Greenstein}(1941)}]{HenyeyApJ1941}
{Henyey}, L.~G. \& {Greenstein}, J.~L. 1941, ApJ, 93, 70

\bibitem[{{Holland} {et~al.}(2006){Holland}, {MacIntosh}, {Fairley}, {Kelly},
  {Montgomery}, {Gostick}, {Atad-Ettedgui}, {Ellis}, {Robson}, {Hollister},
  {Woodcraft}, {Ade}, {Walker}, {Irwin}, {Hilton}, {Duncan}, {Reintsema},
  {Walton}, {Parkes}, {Dunare}, {Fich}, {Kycia}, {Halpern}, {Scott}, {Gibb},
  {Molnar}, {Chapin}, {Bintley}, {Craig}, {Chylek}, {Jenness}, {Economou}, \&
  {Davis}}]{HollandProc2006}
{Holland}, W., {MacIntosh}, M., {Fairley}, A., {et~al.} 2006, in Millimeter and
  Submillimeter Detectors and Instrumentation for Astronomy III., ed.
  J.~{Zmuidzinas} \& W.~D. {Holland} Wayne S.;~Withington, Stafford;~Duncan,
  Vol. 6275

\bibitem[{{Holwerda} {et~al.}(2007){Holwerda}, {Keel}, \&
  {Bolton}}]{HolwerdaAJ2007}
{Holwerda}, B.~W., {Keel}, W.~C., \& {Bolton}, A. 2007, AJ, 134, 2385

\bibitem[{{Ivezic} \& {Elitzur}(1997)}]{IvezicMNRAS1997a}
{Ivezic}, Z. \& {Elitzur}, M. 1997, MNRAS, 287, 799

\bibitem[{{Ivezic} {et~al.}(1997){Ivezic}, {Groenewegen}, {Men'shchikov}, \&
  {Szczerba}}]{IvezicMNRAS1997b}
{Ivezic}, Z., {Groenewegen}, M.~A.~T., {Men'shchikov}, A., \& {Szczerba}, R.
  1997, MNRAS, 291, 121

\bibitem[{{Jarrett} {et~al.}(2003){Jarrett}, {Chester}, {Cutri}, {Schneider},
  \& {Huchra}}]{JarrettAJ2003}
{Jarrett}, T.~H., {Chester}, T., {Cutri}, R., {Schneider}, S.~E., \& {Huchra},
  J.~P. 2003, AJ, 125, 525

\bibitem[{{Jonsson}(2006)}]{JonssonMNRAS2006}
{Jonsson}, P. 2006, MNRAS, 372, 2

\bibitem[{{Juvela}(2005)}]{JuvelaA&A2005}
{Juvela}, M. 2005, A\&A, 440, 531

\bibitem[{{Kennicutt}(1998)}]{KennicuttARA&A1998}
{Kennicutt}, R.~C. 1998, ARA\&A, 36, 189

\bibitem[{{Kreysa} {et~al.}(2003){Kreysa}, {Bertoldi}, {Gemuend}, {Menten},
  {Muders}, {Reichertz}, {Schilke}, {Chini}, {Lemke}, {May}, {Meyer}, \&
  {Zakosarenko}}]{KreysaProc2003}
{Kreysa}, E., {Bertoldi}, F., {Gemuend}, H.-P., {et~al.} 2003, in Millimeter
  and Submillimeter Detectors for Astronomy., ed. T.~G. {Phillips} \&
  J.~{Zmuidzinas}, Vol. 4855, 41--48

\bibitem[{{Kurosawa} \& {Hillier}(2001)}]{KurosawaA&A2001}
{Kurosawa}, R. \& {Hillier}, D.~J. 2001, A\&A, 379, 336

\bibitem[{{Kylafis} \& {Bahcall}(1987)}]{KylafisApJ1987}
{Kylafis}, N.~D. \& {Bahcall}, J.~N. 1987, ApJ, 317, 637

\bibitem[{{Leitherer} {et~al.}(1996){Leitherer}, {Alloin}, {Fritz-V.
  Alvensleben}, {Gallagher}, {Huchra}, {Matteucci}, {O'Connell}, {Beckman},
  {Bertelli}, {Bica}, {Boisson}, {Bonatto}, {Bothun}, {Bressan}, {Brodie},
  {Bruzual}, {Burstein}, {Buser}, {Caldwell}, {Casuso}, {Cervino}, {Charlot},
  {Chavez}, {Chiosi}, {Christian}, {Cuisinier}, {Dallier}, {De Koter},
  {Delisle}, {Diaz}, {Dopita}, {Dorman}, {Fagotto}, {Fanelli}, {Fioc},
  {Garcia-Vargas}, {Girardi}, {Goldader}, {Hardy}, {Heckman}, {Iglesias},
  {Jablonka}, {Joly}, {Jones}, {Kurth}, {Lancon}, {Lejeune}, {Loxen}, {Maeder},
  {Malagnini}, {Marigo}, {Mas-Hesse}, {Meynet}, {Moller}, {Molla}, {Morossi},
  {Nasi}, {Nichols}, {Odegaard}, {Parker}, {Pastoriza}, {Peletier}, {Robert},
  {Rocca-Volmerange}, {Schaerer}, {Schmidt}, {Schmitt}, {Schommer}, {Schmutz},
  {Roos}, {Silva}, {Stasinska}, {Sutherland}, {Tantalo}, {Traat}, {Vallenari},
  {Vazdekis}, {Walborn}, {Worthey}, \& {Wu}}]{LeithererPASP1996}
{Leitherer}, C., {Alloin}, D., {Fritz-V. Alvensleben}, U., {et~al.} 1996, PASP,
  108, 996

\bibitem[{{Li} \& {Draine}(2001)}]{LiApJ2001}
{Li}, A. \& {Draine}, B.~T. 2001, ApJ, 554, 778

\bibitem[{{Li} {et~al.}(2008){Li}, {Hopkins}, {Hernquist}, {Finkbeiner}, {Cox},
  {Springel}, {Jiang}, {Fan}, \& {Yoshida}}]{LiApJ2008}
{Li}, Y., {Hopkins}, P.~F., {Hernquist}, L., {et~al.} 2008, ApJ, 678, 41

\bibitem[{{Lucy}(1999)}]{LucyA&A1999}
{Lucy}, L.~B. 1999, A\&A, 344, 282

\bibitem[{{Mathis} {et~al.}(1983){Mathis}, {Mezger}, \&
  {Panagia}}]{MathisA&A1983}
{Mathis}, J.~S., {Mezger}, P.~G., \& {Panagia}, N. 1983, A\&A, 128, 212

\bibitem[{{Matthews} \& {Wood}(2001)}]{MatthewsApJ2001}
{Matthews}, L.~D. \& {Wood}, K. 2001, ApJ, 548, 150

\bibitem[{{Mie}(1908)}]{MieAnnPhys1908}
{Mie}, G. 1908, Ann. Phys., 25, 377

\bibitem[{{Misiriotis} \& {Bianchi}(2002)}]{MisiriotisA&A2002}
{Misiriotis}, A. \& {Bianchi}, S. 2002, A\&A, 384, 866

\bibitem[{{Misiriotis} {et~al.}(2001){Misiriotis}, {Popescu}, {Tuffs}, \&
  {Kylafis}}]{MisiriotisA&A2001}
{Misiriotis}, A., {Popescu}, C.~C., {Tuffs}, R., \& {Kylafis}, N.~D. 2001,
  A\&A, 372, 775

\bibitem[{{Misselt} {et~al.}(2001){Misselt}, {Gordon}, {Clayton}, \&
  {Wolff}}]{MisseltApJ2001}
{Misselt}, K.~A., {Gordon}, K.~D., {Clayton}, G.~C., \& {Wolff}, M.~J. 2001,
  ApJ, 551, 277

\bibitem[{{Niccolini} {et~al.}(2003){Niccolini}, {Woitke}, \&
  {Lopez}}]{NiccoliniA&A2003}
{Niccolini}, G., {Woitke}, P., \& {Lopez}, B. 2003, A\&A, 399, 703

\bibitem[{{Oosterloo} {et~al.}(2007){Oosterloo}, {Fraternali}, \&
  {Sancisi}}]{OosterlooAJ2007}
{Oosterloo}, T., {Fraternali}, F., \& {Sancisi}, R. 2007, AJ, 134, 1019

\bibitem[{{Pascucci} {et~al.}(2004){Pascucci}, {Wolf}, {Steinacker},
  {Dullemond}, {Henning}, {Niccolini}, {Woitke}, \& {Lopez}}]{PascucciA&A2004}
{Pascucci}, I., {Wolf}, S., {Steinacker}, J., {et~al.} 2004, A\&A, 417, 793

\bibitem[{{Pierini} {et~al.}(2004){Pierini}, {Gordon}, {Witt}, \&
  {Madsen}}]{PieriniApJ2004}
{Pierini}, D., {Gordon}, K.~D., {Witt}, A.~N., \& {Madsen}, G.~J. 2004, ApJ,
  617, 1022

\bibitem[{{Popescu} {et~al.}(2000){Popescu}, {Misiriotis}, {Kylafis}, {Tuffs},
  \& {Fischera}}]{PopescuA&A2000}
{Popescu}, C.~C., {Misiriotis}, A., {Kylafis}, N.~D., {Tuffs}, R.~J., \&
  {Fischera}, J. 2000, A\&A, 362, 138

\bibitem[{{Popescu} {et~al.}(2004){Popescu}, {Tuffs}, {Kylafis}, \&
  {Madore}}]{PopescuA&A2003a}
{Popescu}, C.~C., {Tuffs}, R.~J., {Kylafis}, N.~D., \& {Madore}, B.~F. 2004,
  A\&A, 414, 45

\bibitem[{{Prugniel} \& {Simien}(1997)}]{PrugnielA&A1997}
{Prugniel}, P. \& {Simien}, F. 1997, A\&A, 321, 111

\bibitem[{{Regan} {et~al.}(2001){Regan}, {Thornley}, {Helfer}, {Sheth}, {Wong},
  {Vogel}, {Blitz}, \& {Bock}}]{ReganApJ2001}
{Regan}, M.~W., {Thornley}, M.~D., {Helfer}, T.~T., {et~al.} 2001, ApJ, 561,
  218

\bibitem[{{Rieke} {et~al.}(2004){Rieke}, {Young}, {Engelbracht}, {Kelly},
  {Low}, {Haller}, {Beeman}, {Gordon}, {Stansberry}, {Misselt}, {Cadien},
  {Morrison}, {Rivlis}, {Latter}, {Noriega-Crespo}, {Padgett}, {Stapelfeldt},
  {Hines}, {Egami}, {Muzerolle}, {Alonso-Herrero}, {Blaylock}, {Dole}, {Hinz},
  {Le Floc'h}, {Papovich}, {P{\'e}rez-Gonz{\'a}lez}, {Smith}, {Su}, {Bennett},
  {Frayer}, {Henderson}, {Lu}, {Masci}, {Pesenson}, {Rebull}, {Rho}, {Keene},
  {Stolovy}, {Wachter}, {Wheaton}, {Werner}, \& {Richards}}]{RiekeApJS2004}
{Rieke}, G.~H., {Young}, E.~T., {Engelbracht}, C.~W., {et~al.} 2004, ApJS, 154,
  25

\bibitem[{{Rocha} {et~al.}(2008){Rocha}, {Jonsson}, {Primack}, \&
  {Cox}}]{RochaMNRAS2008}
{Rocha}, M., {Jonsson}, P., {Primack}, J.~R., \& {Cox}, T.~J. 2008, MNRAS, 383,
  1281

\bibitem[{{Rosolowsky}(2005)}]{RosolowskyPASP2005}
{Rosolowsky}, E. 2005, PASP, 117, 1403

\bibitem[{{Rosolowsky} {et~al.}(2003){Rosolowsky}, {Engargiola}, {Plambeck}, \&
  {Blitz}}]{RosolowskyApJ2003}
{Rosolowsky}, E., {Engargiola}, G., {Plambeck}, R., \& {Blitz}, L. 2003, ApJ,
  599, 258

\bibitem[{{Sanders} {et~al.}(2003){Sanders}, {Mazzarella}, {Kim}, {Surace}, \&
  {Soifer}}]{SandersAJ2003}
{Sanders}, D.~B., {Mazzarella}, J.~M., {Kim}, D.-C., {Surace}, J.~A., \&
  {Soifer}, B.~T. 2003, AJ, 126, 1607

\bibitem[{{Sersic}(1968)}]{SersicBook1968}
{Sersic}, J.~L. 1968, {Atlas de galaxias australes} (Cordoba, Argentina:
  Observatorio Astronomico, 1968)

\bibitem[{{Siewert} \& {Maiorino}(1979)}]{SiewertJQSRT1979}
{Siewert}, C.~E. \& {Maiorino}, J.~R. 1979, Journal of Quantitative
  Spectroscopy and Radiative Transfer, 22, 435

\bibitem[{{Silva} {et~al.}(1998){Silva}, {Granato}, {Bressan}, \&
  {Danese}}]{SilvaApJprep1998}
{Silva}, L., {Granato}, G.~L., {Bressan}, A., \& {Danese}, L. 1998, ApJ, 509,
  103

\bibitem[{{Sofue} \& {Nakai}(1993)}]{SofuePASJ1993}
{Sofue}, Y. \& {Nakai}, N. 1993, PASJ, 45, 139

\bibitem[{{Tuffs} {et~al.}(2004){Tuffs}, {Popescu}, {V{\" o}lk}, {Kylafis}, \&
  {Dopita}}]{TuffsA&A2004}
{Tuffs}, R.~J., {Popescu}, C.~C., {V{\" o}lk}, H.~J., {Kylafis}, N.~D., \&
  {Dopita}, M.~A. 2004, A\&A, 419, 821

\bibitem[{{V{\'a}rosi} \& {Dwek}(1999)}]{VarosiPrep1999}
{V{\'a}rosi}, F. \& {Dwek}, E. 1999, ApJ, 523, 265

\bibitem[{{Weingartner} \& {Draine}(2001{\natexlab{a}})}]{WeingartnerApJ2001a}
{Weingartner}, J.~C. \& {Draine}, B.~T. 2001{\natexlab{a}}, ApJ, 548, 296

\bibitem[{{Weingartner} \& {Draine}(2001{\natexlab{b}})}]{WeingartnerApJS2001}
{Weingartner}, J.~C. \& {Draine}, B.~T. 2001{\natexlab{b}}, ApJS, 134, 263

\bibitem[{{Witt} \& {Gordon}(1996)}]{WittApJ1996}
{Witt}, A.~N. \& {Gordon}, K.~D. 1996, ApJ, 463, 681

\bibitem[{{Wolf}(2003)}]{WolfApJ2003}
{Wolf}, S. 2003, ApJ, 582, 859

\bibitem[{{Wolf} {et~al.}(1999){Wolf}, {Henning}, \& {Stecklum}}]{WolfA&A1999}
{Wolf}, S., {Henning}, T., \& {Stecklum}, B. 1999, A\&A, 349, 839

\bibitem[{{Xilouris} {et~al.}(1998){Xilouris}, {Alton}, {Davies}, {Kylafis},
  {Papamastorakis}, \& {Trewhella}}]{XilourisA&A1998}
{Xilouris}, E.~M., {Alton}, P.~B., {Davies}, J.~I., {et~al.} 1998, A\&A, 331,
  894

\bibitem[{{Xilouris} {et~al.}(1999){Xilouris}, {Byun}, {Kylafis}, {Paleologou},
  \& {Papamastorakis}}]{XilourisSub1998}
{Xilouris}, E.~M., {Byun}, Y.~I., {Kylafis}, N.~D., {Paleologou}, E.~V., \&
  {Papamastorakis}, J. 1999, A\&A, 344, 868

\bibitem[{{Young} \& {Scoville}(1991)}]{YoungARA&A1991}
{Young}, J.~S. \& {Scoville}, N.~Z. 1991, ARA\&A, 29, 581

\bibitem[{{Yusef-Zadeh} {et~al.}(1984){Yusef-Zadeh}, {Morris}, \&
  {White}}]{YusefZadehApJ1984}
{Yusef-Zadeh}, F., {Morris}, M., \& {White}, R.~L. 1984, ApJ, 278, 186

\end{thebibliography}

\end{document}